\documentclass[lettersize,journal]{IEEEtran}
\usepackage{fix-cm}

\usepackage{cite}
\usepackage{amsmath,amssymb,amsfonts}
\usepackage{amsthm}

\theoremstyle{remark}

\usepackage{graphicx}
\usepackage{xcolor}
\usepackage{textcomp}
\usepackage{multirow}
\usepackage{booktabs}
\usepackage{array}
\usepackage{stfloats}
\usepackage{url}
\usepackage{verbatim}
\usepackage{capt-of}
\usepackage[caption=false,font=normalsize,labelfont=sf,textfont=sf]{subfig}

\usepackage{algorithm}
\usepackage[noend]{algpseudocode}

\usepackage[hidelinks]{hyperref}

\usepackage{bbm}

\graphicspath{{FIGURES/}}

\begin{document}

\title{Integrated Sensing, Communication, and Computing for NR-V2X: A Cross-Layer Resource Allocation Framework Using Multi-Agent Reinforcement Learning}


\author{Indulekha K.~P., \IEEEmembership{Student Member, IEEE}, 
and T.~G. Venkatesh, \IEEEmembership{Member, IEEE}%

\thanks{Indulekha K.~P. and T.~G. Venkatesh are with the Department of Electrical Engineering, Indian Institute of Technology Madras, Chennai 600036, India 
(e-mail: ee22d006@smail.iitm.ac.in; tgvenky@ee.iitm.ac.in).}%
}



\maketitle

\begin{abstract}
Integrated sensing, communication, and computation (ISCC) is emerging as a unified design paradigm for future vehicular networks that require joint environment perception, safety-critical information exchange, and latency-sensitive task processing. In New Radio Vehicle-to-Everything (NR-V2X) Mode~2, autonomous resource selection is performed through sensing-based semi-persistent scheduling (SB-SPS), which is effective for distributed communication resource reservation but does not explicitly consider sensing-resource demand, task-induced computation workload, and the additional latency introduced by mobile edge computing (MEC) offloading. This paper develops multi-agent proximal policy optimization-based SB-SPS (MAPPO-SPS), an ISCC-aware cross-layer scheduler that jointly adapts SB-SPS reservation, radio-resource partitioning, and overflow-driven computation-offloading decisions at control epochs. The scheduling problem is formulated as a cooperative partially observable Markov game and solved using MAPPO with centralized training and decentralized execution (CTDE). Simulation results show that MAPPO-SPS achieves a balanced tradeoff among CRLB-based sensing accuracy, packet reception ratio (PRR), effective throughput, energy consumption, and end-to-end delay.
\end{abstract}

\begin{IEEEkeywords}
New Radio Vehicle-to-Everything (NR-V2X), integrated sensing, communication, and computation (ISCC), sensing-based semi-persistent scheduling (SB-SPS), cross-layer scheduling, multi-agent reinforcement learning, multi-agent proximal policy optimization (MAPPO), mobile edge computing (MEC), computation offloading.
\end{IEEEkeywords}

\section{Introduction}
\IEEEpeerreviewmaketitle
\IEEEPARstart{T}{he} growing demand for reliable vehicular connectivity has made Vehicle-to-Everything (V2X) communications a core enabler of connected and automated driving. Standardized by the Third Generation Partnership Project (3GPP), V2X has evolved from Long Term Evolution Vehicle-to-Everything (LTE-V2X) to Fifth Generation New Radio Vehicle-to-Everything (5G NR-V2X) to meet stringent latency, reliability, and mobility requirements~\cite{Garcia_CST_2021}. Resource allocation in Cellular Vehicle-to-Everything (C-V2X) is defined under two operational modes, namely a centralized under-coverage mode (Mode~3 in LTE-V2X and Mode~1 in NR-V2X) and a distributed out-of-coverage mode (Mode~4 in LTE-V2X and Mode~2 in NR-V2X), where the latter relies on sensing-based semi-persistent scheduling (SPS) for autonomous resource selection \cite{Sehla_IoTJ_2022}. The performance of 5G NR-V2X sidelink Mode~2 has been extensively analyzed using open-source simulation platforms to evaluate key metrics such as reliability, latency, and resource utilization \cite{Todisco_ACCESS_2021}.

\subsection{Literature Review}
Modern intelligent vehicles are equipped with multiple sensing modalities, such as cameras, LiDAR, radar, and inertial sensors, which generate massive volumes of multi-dimensional perception data that require timely processing and management for reliable autonomous operation \cite{Choi_CommMag_2016}. Integrated Sensing and Communication (ISAC) has emerged as a key paradigm for future wireless systems, enabling the joint utilization of spectrum and hardware to simultaneously support sensing and communication \cite{Liu_CST_2022}. Recent studies have further highlighted that large-scale networked sensing generates massive perception data requiring reliable transmission and real-time processing, thereby motivating a tighter integration of sensing, communication, and computation resources \cite{Lu_IoTJ_2024}. In the V2X context, recent works have already explored parts of this coupled design space. For instance, Long \emph{et al.} employed deep reinforcement learning for joint RIS beamforming and ISAC parameter configuration under imperfect channels \cite{Long2024DRL}, while He \emph{et al.} proposed a full-duplex ISAC framework that jointly optimizes transmit beamforming and power allocation under uplink--downlink interference constraints \cite{He2023FDISAC}.

The fundamental trade-off between communication overhead and computational latency in resource-constrained intelligent systems has also been highlighted in prior work \cite{Shao_CommMag_2020}, emphasizing the need to jointly manage transmission and processing resources in distributed edge intelligence. In this direction, MEC-oriented studies have shown that task partitioning, transmission power, and CPU-frequency adaptation must be jointly optimized in order to satisfy latency and energy constraints \cite{Cao_WiOpt_2018,Cao_TWC_2020}. More broadly, recent sensing-driven edge intelligence and mobile sensing studies have further indicated that large volumes of sensing data generated by distributed devices require efficient communication and coordinated computing support for timely processing \cite{Capponi_COMST_2019,Cai_TNSE_2022}. These observations suggest that communication and computation can no longer be designed independently once sensing-intensive vehicular services are considered.

Machine learning (ML)-based resource management has become increasingly relevant in V2X systems because network conditions vary rapidly with vehicle mobility, traffic density, and interference. Reinforcement learning (RL) is particularly attractive in this setting, as it can learn adaptive resource-allocation policies directly from experience in dynamic environments \cite{Tang_CST_2021}. Moreover, the inherently distributed and highly dynamic nature of V2X networks, characterized by rapid topology changes, mutual interference, and coupled resource decisions among vehicles, naturally motivates multi-agent reinforcement learning (MARL) for scalable and adaptive decentralized decision-making \cite{Althamary_WCMC_2019}. In this direction, Gu \emph{et al.} developed a multi-agent reinforcement learning-based SPS scheme for C-V2X Mode~4, showing that the resource reselection problem can be reformulated as a Markov game and addressed in a decentralized manner using local observations \cite{Gu2022RLSPS}. These developments further indicate that, as intelligent vehicular services become more data-intensive and delay-sensitive, sensing, communication, and computation can no longer be treated as isolated functions. Instead, they must be coordinated within a unified cross-layer design, which naturally motivates Integrated Sensing, Communication, and Computation (ISCC) as a promising framework for joint resource allocation and end-to-end performance optimization \cite{Wen2024Survey}. Motivated by this tight cross-layer coupling, we aim to develop a reinforcement learning-based ISCC-aware MAC framework for next-generation NR-V2X systems that jointly optimizes radio resource allocation and computation offloading for sensing and communication tasks.

\section{Related Works and Contributions}

To further position the proposed framework, we next focus on the works most closely related to ISCC-aware vehicular resource allocation. Qi \textit{et al.} proposed a unified ISCC design framework for 6G networks, formulating multi-objective optimization problems that jointly optimize sensing, computing, and communication through beamforming and power allocation under transmit power and QoS constraints \cite{Qi2022}. Du \textit{et al.} investigated an ISCC-enabled over-the-air federated learning framework in which sensing, communication, and computation share common time--frequency resources and are jointly optimized through transceiver beamforming and device selection \cite{du2024iscc_federated}. These studies clearly demonstrate the value of joint cross-domain optimization, but they are primarily developed for centralized systems with controller-driven coordination.

A similar direction has been explored in MEC-enabled and ISAC-assisted resource-allocation works. Liang \textit{et al.} developed a two-level decomposition framework for joint task offloading and communication--computation allocation under continuous spectrum sharing \cite{Liang2024}. Liu \textit{et al.} studied ISAC-aided V2X resource allocation using Lyapunov optimization and MDP-based control for joint computation offloading and wireless resource management \cite{liu2023isac_v2x}. Huang \textit{et al.} investigated sensing-aware communication--computation optimization with IRS assistance under energy and latency constraints \cite{Huang2022ISAC}. Zhao \textit{et al.} formulated ISCC radio resource allocation with network externalities and proposed a matching-based scheduler for device association and subchannel assignment \cite{Zhao2022RRAlloc}. He \textit{et al.} modeled competition among sensing, communication, and computation tasks at an edge-enabled base station and designed a joint resource-allocation and action-detection framework \cite{he2024iscc_framework}. Although these works capture important parts of sensing--communication--computation interaction, they still rely on centralized optimization or infrastructure-assisted scheduling and therefore do not reflect the autonomous sidelink contention and reservation dynamics of NR-V2X Mode~2.

Among the closest V2X-oriented formulations, Liu \textit{et al.}~\cite{Liu_TCOM_2025_ISCC} proposed a joint radio and computation resource-allocation framework for ISCC-enabled vehicular networks. Their work is highly relevant in cross-layer spirit, but it is built on deterministic centralized optimization with global system information and does not incorporate decentralized MAC-layer decision making or learning-based adaptation. Consequently, the joint integration of sensing quality, communication reliability, and computation offloading within the decentralized NR-V2X Mode~2 sensing-based semi-persistent scheduling (SB-SPS) framework remains insufficiently explored.

This gap is particularly important because, in practical NR-V2X Mode~2 systems, sensing, communication, and computation contend for limited radio and computational resources, whereas the SB-SPS resource-allocation mechanism is still primarily communication-centric and does not explicitly account for sensing-workload generation or computation queue states associated with perception processing and task offloading \cite{Sehla_IoTJ_2022}. Motivated by these limitations, we develop a unified cross-layer ISCC framework for NR-V2X Mode~2 that integrates sensing, communication, and computation within the SB-SPS access process under decentralized operation.

The main contributions of this work are summarized as follows:

\begin{itemize}
\item We extend NR-V2X Mode~2 SB-SPS into an ISCC-aware cross-layer reservation framework that partitions PC5 sidelink resources for sensing and communication while modeling V2I offloading separately.

\item We develop a coupled sensing--communication--computation model that captures the tradeoff among sensing accuracy, communication efficiency, and computation cost under shared radio and MEC resources.

\item We model the decentralized ISCC-aware scheduling problem as a cooperative partially observable Markov game and solve it using MAPPO-SPS under the CTDE paradigm.

\item We evaluate MAPPO-SPS against MA-A2C-SPS and two greedy extended SB-SPS baselines under identical Mode~2 settings across sensing, communication, and computation metrics.
\end{itemize}

\section{System Model}
\label{sec:system_model}

\begin{figure}[!ht]
    \centering
    \includegraphics[height=6cm, width=9cm]{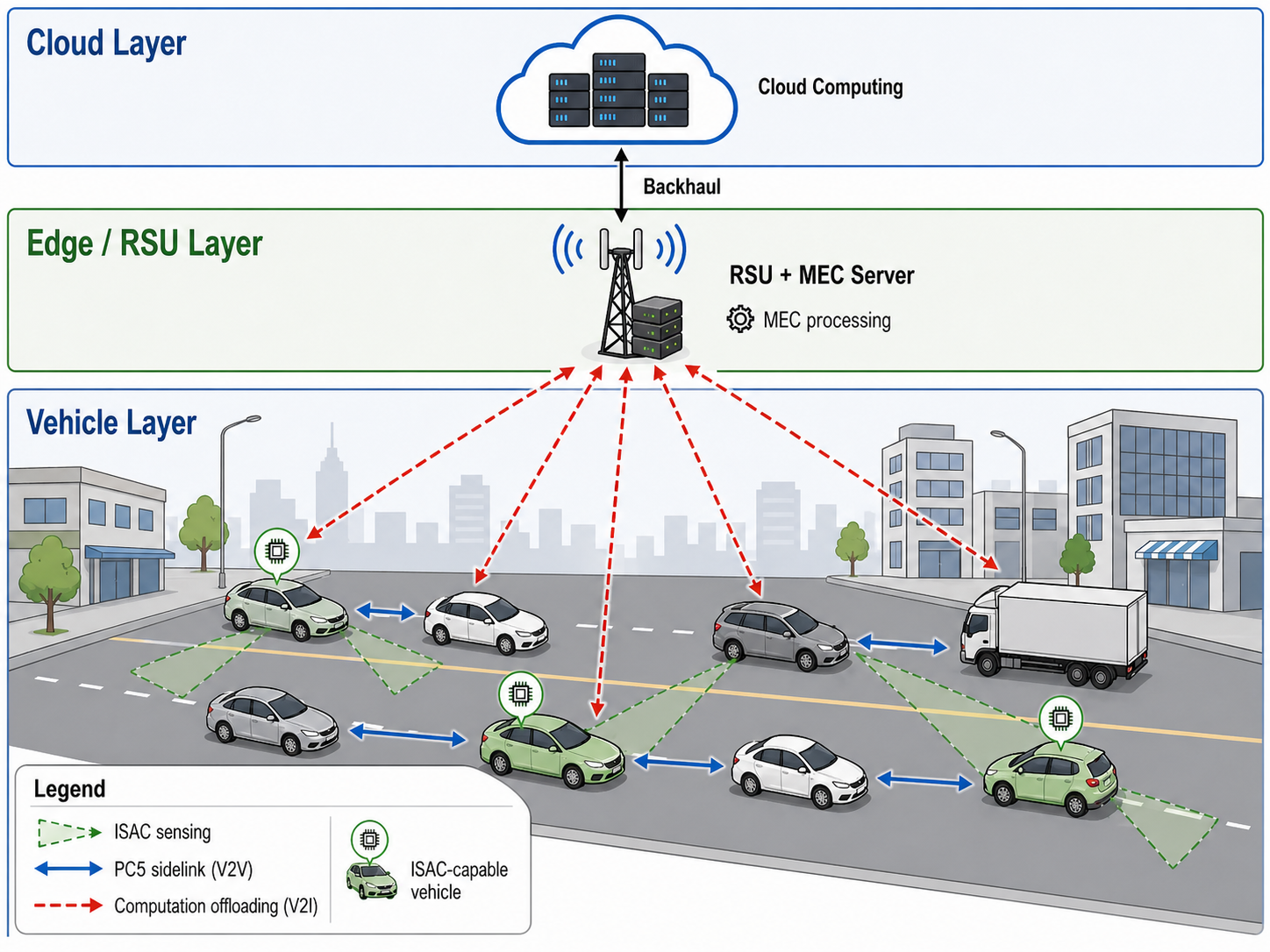}
    \caption{NR-V2X Mode-2 ISCC architecture with joint sensing, sidelink communication, and MEC-assisted computation.}
    \label{fig:system_architecture}
\end{figure}

We consider an NR-V2X sidelink network operating under 3GPP Release~16 Mode~2, with \(N\) vehicles
\(\mathcal{U}=\{u_1,u_2,\ldots,u_N\}\) moving along a road segment, as shown in Fig.~\ref{fig:system_architecture}. A single MEC-enabled RSU provides roadside computation support, and all vehicles are assumed to remain within its coverage during the scheduling horizon. This avoids inter-RSU association and handover modeling, allowing the focus to remain on MAC-layer ISCC scheduling. Vehicles exchange safety messages over the PC5 sidelink, while computation overflow is offloaded to the MEC server through a V2I uplink. The V2I channel follows a block-fading model, remaining quasi-static within each slot and varying independently across slots. Each vehicle is equipped with an OFDM-based ISAC transceiver and an on-board processor with finite local CPU capacity \(C_i\) cycles/slot. The main notations are summarized in Table~\ref{tab:main_notations}. In NR-V2X, Mode~1 relies on network-controlled resource allocation, whereas Mode~2 allows vehicles to autonomously select sidelink resources. This work focuses on Mode~2 with sensing-based semi-persistent scheduling (SB-SPS). Under SB-SPS, a vehicle senses the channel over a sensing window, identifies candidate resources within a selection window, and reserves resources for future transmissions separated by the resource reservation interval (RRI). The reservation is maintained according to the reselection counter (RC), while occupied resources are excluded using sidelink control information and received-signal measurements.

Time is divided into \(T\) NR slots indexed by \(t\in\{1,\ldots,T\}\). Each slot contains 14 OFDM symbols with subcarrier spacing \(\Delta f\), yielding slot duration \(T_{\mathrm{slot}}=2^{-\mu}\) ms, where \(\mu\) is the NR numerology index. The sidelink resource pool contains \(N_{\mathrm{PRB}}\) PRBs, each comprising 12 contiguous subcarriers. We adopt a two-timescale model: packet arrivals, sensing-data generation, channel realizations, queue evolution, local computation, and transmission outcomes evolve at the slot level, whereas SB-SPS reservation updates and MAPPO actions occur at control epochs indexed by \(k\). Let \(\mathcal{T}_k\) denote the set of slots associated with epoch \(k\). At epoch \(k\), vehicle \(u_i\) determines its sidelink reservation, SB-SPS persistence parameters, and cross-layer resource split, which remain fixed for all \(t\in\mathcal{T}_k\) unless re-evaluation or reselection is triggered.

\begin{table}[!ht]
\centering
\caption{Key Notations Used for the ISCC Framework}
\label{tab:main_notations}
\renewcommand{\arraystretch}{1.05}
\begin{tabular}{p{0.25\linewidth} p{0.67\linewidth}}
\hline
\textbf{Notation} & \textbf{Description} \\
\hline
\(\mathcal{U}\) & Set of vehicles/agents, \(\mathcal{U}=\{u_1,\ldots,u_N\}\) \\
\(N\) & Number of vehicles in the NR-V2X network \\
\(t,k\) & NR slot index and SB-SPS control epoch index \\
\(\mathcal{T}_k\) & Set of NR slots associated with epoch \(k\) \\
\(T_{\mathrm{slot}},T_{\mathrm{sym}}\) & NR slot duration and useful OFDM symbol duration \\
\(\Delta f\) & OFDM subcarrier spacing \\
\(N_{\mathrm{PRB}}\) & Number of PRBs in the sidelink resource pool \\
\(N_i^{\mathrm{sl}}(k)\) & PC5 sidelink PRBs reserved by vehicle \(u_i\) \\
\(N_i^s(k),N_i^c(k)\) & Sensing and communication PRBs on PC5 sidelink \\
\(N_i^o(k)\) & V2I uplink PRBs used for computation offloading \\
\(N_i^{o,\max}(k)\) & Maximum V2I uplink PRBs available for offloading \\
\(M_i^s(k)\) & OFDM symbols assigned to sensing \\
\(r_i(k)\) & Selected SB-SPS reservation resource \\
\(RC_i(k)\) & Reselection counter selected at SB-SPS update \\
\(P_{\mathrm{keep},i}(k)\) & Resource keep probability \\
\(\mathcal{R}_i^{\mathrm{cand}}(k)\) & Candidate reservation-resource set \\
\(\mathcal C_{RC}(T_{\mathrm{RRI}})\) & RRI-dependent admissible reselection-counter set \\
\(B_i^s(k),B_i^c(k)\) & Sensing and communication bandwidths\\
\(\gamma_{i,\ell}^{s}(t)\) & Sensing SNR for target \(\ell\) observed by vehicle \(u_i\) \\
\(\sigma_{\ell}^{\mathrm{RCS}}(t)\) & Radar cross section of target \(\ell\) \\
\(\rho_{\mathrm{SI}}\) & Residual self-interference coefficient after cancellation \\
\(\mathcal T_i(t)\) & Detected target set of vehicle \(u_i\) \\
\(\epsilon_i^{\mathrm{sens}}(t)\) & CRLB-based sensing penalty of vehicle \(u_i\) \\
\(\widetilde{\epsilon}_i^{\mathrm{sens}},\widetilde{\Phi}_i^c,\widetilde{\psi}_i^{\mathrm{comp}}\) & Normalized sensing, communication-deficiency, and computation costs \\
\(R_i^c(t),R_i^{\mathrm{eff}}(t)\) & Sidelink rate and effective sidelink rate \\
\(A_i^{\mathrm{comm}}(t)\) & Communication traffic arrival at vehicle \(u_i\) \\
\(Q_i^{\mathrm{comm}}(t)\) & Communication queue length of vehicle \(u_i\) \\
\(V_i^s(t),V_i^c(t)\) & Sensing and communication computation workloads \\
\(V_i^{\mathrm{ov},s}(t),V_i^{\mathrm{ov},c}(t)\) & Sensing and communication overflow workloads \\
\(C_i,C_{\mathrm R}\) & Local CPU capacity and MEC server processing capacity \\
\(L_{\mathrm R}^{s}(t),L_{\mathrm R}^{c}(t)\) & MEC sensing and communication queue backlogs \\
\(\eta_i^c(k),\eta_i^s(k)\) & Overflow-decision variables \\
\(\ell_i(t)\) & Normalized cross-layer cost of vehicle \(u_i\) \\
\hline
\end{tabular}
\end{table}

Let \(N_i^{\mathrm{sl}}(k)\) denote the number of PC5 sidelink PRBs reserved by vehicle \(u_i\) at epoch \(k\). These PRBs are divided between sensing and communication such that \(N_i^s(k)+N_i^c(k)\leq N_i^{\mathrm{sl}}(k)\), where \(N_i^s(k)\) and \(N_i^c(k)\) are the PRBs used for OFDM-based sensing and sidelink communication, respectively. This partition extends the SB-SPS reservation structure to ISCC operation without changing the decentralized Mode~2 sensing-and-selection principle. In addition, vehicle \(u_i\) may use V2I uplink resources for MEC offloading. Since PC5 sidelink and V2I uplink are distinct links, the offloading resource is modeled separately from the SB-SPS-controlled sidelink reservation. Let \(N_i^o(k)\) denote the number of V2I uplink PRBs assigned to vehicle \(u_i\), with \(0\le N_i^o(k)\le N_i^{o,\max}(k)\). Hence, \(N_i^o(k)\) is not included in the PC5 sidelink partition. Increasing \(N_i^s(k)\) improves sensing accuracy but increases sensing workload; increasing \(N_i^c(k)\) improves sidelink service; and increasing \(N_i^o(k)\) improves overflow delivery to the MEC server but may increase V2I transmission and queueing delay.

\subsection{Sensing Model}
\label{subsec:sensing}

Each vehicle \(u_i\) is equipped with an OFDM-based ISAC transceiver, where OFDM symbols can be reused for radar-type delay-Doppler sensing~\cite{Sturm2011_OFDMRadar,Decarli2024JCSNRV2X}. During slot \(t\in\mathcal{T}_k\), vehicle \(u_i\) performs sensing using the sensing PRBs and OFDM symbols selected at control epoch \(k\). The sensing bandwidth is
\begin{equation}
    B_i^s(k)=12N_i^s(k)\Delta f,
    \label{eq:sensing_bw}
\end{equation}
where \(N_i^s(k)\) is the number of sensing PRBs and \(\Delta f\) is the subcarrier spacing. Let
\[
N_{s,i}(k)\triangleq 12N_i^s(k), \qquad M_{s,i}(k)\triangleq M_i^s(k),
\]
where \(M_i^s(k)\in\{1,\ldots,14\}\) is the number of OFDM symbols assigned to sensing. The subcarrier and OFDM-symbol indices within the sensing block are denoted by
\(n\in\{0,\ldots,N_{s,i}(k)-1\}\) and \(m\in\{0,\ldots,M_{s,i}(k)-1\}\), respectively.

\subsubsection{Signal Model}

Let \(\tilde{x}_i[n,m]\) denote the transmitted OFDM sensing symbol of vehicle \(u_i\) on subcarrier \(n\) and OFDM symbol \(m\), with average power \(\mathbb{E}[|\tilde{x}_i[n,m]|^2]=p_i^{\mathrm{tx}}/N_{s,i}(k)\), where \(p_i^{\mathrm{tx}}\) is the sensing transmit power. The transmitted sensing symbols are assumed known at the receiver. For target \(\ell\in\{1,\ldots,\mathcal P\}\), let \(r_\ell(t)\) and \(v_\ell(t)\) denote its range and radial velocity in slot \(t\), respectively. The corresponding round-trip delay and Doppler shift are
\begin{equation}
\begin{aligned}
    \tau_\ell(t) &= \frac{2r_\ell(t)}{c_0}, &
    \nu_\ell(t) &= \frac{2v_\ell(t)}{\lambda_c},
\end{aligned}
\label{eq:delay_doppler}
\end{equation}
where \(c_0\) is the speed of light and \(\lambda_c=c_0/f_c\) is the carrier wavelength.

After down-conversion and cyclic-prefix removal, the received sensing echo is modeled as
\begin{multline}
    y_i[n,m,t]
    =
    \sum_{\ell=1}^{\mathcal P}
    \zeta_\ell(t)\tilde{x}_i[n,m]
    e^{-j2\pi n\Delta f\tau_\ell(t)}
    e^{j2\pi \nu_\ell(t)mT_{\mathrm{sym}}}
    \\
    + z_i[n,m,t],
    \label{eq:echo}
\end{multline}
where \(\zeta_\ell(t)\in\mathbb C\) is the complex reflection coefficient of target \(\ell\), \(T_{\mathrm{sym}}\approx1/\Delta f\) is the useful OFDM symbol duration excluding the cyclic prefix, and
\(z_i[n,m,t]\sim\mathcal{CN}(0,\sigma_z^2)\) denotes sensing noise.

Since the transmitted OFDM symbols are known, the normalized sensing samples are
\begin{align}
\tilde y_i[n,m,t]
&= \frac{y_i[n,m,t]}{\tilde x_i[n,m]} \notag\\
&=
\sum_{\ell=1}^{\mathcal P}
\zeta_\ell(t)
e^{-j2\pi n\Delta f\tau_\ell(t)}
e^{j2\pi \nu_\ell(t)mT_{\mathrm{sym}}}
+\tilde z_i[n,m,t],
\label{eq:channel_est}
\end{align}
where \(\tilde z_i[n,m,t]\) denotes the post-normalization noise, whose variance is absorbed into the effective sensing SNR.

\subsubsection{Range-Doppler Processing and Workload}

A two-dimensional DFT over the sensing subcarriers and OFDM symbols produces the range-Doppler map
\begin{equation}
\begin{aligned}
\mathcal P_i[p,q,t]
=
\left|
\sum_{n=0}^{N_{s,i}(k)-1}
\sum_{m=0}^{M_{s,i}(k)-1}
\tilde y_i[n,m,t]\,
\psi_n(p)\,\varphi_m(q)
\right|^2 ,
\end{aligned}
\label{eq:rdmap}
\end{equation}
where \(p\) and \(q\) denote the range and Doppler bin indices, respectively.

The sensing data generated by vehicle \(u_i\) in slot \(t\in\mathcal{T}_k\) is modeled as
\begin{equation}
    D_i^s(t)=2N_{s,i}(k)M_{s,i}(k)b,
    \label{eq:sensing_data}
\end{equation}
where \(D_i^s(t)\) is measured in bits and \(b\) is the number of quantization bits per real-valued I/Q component. The corresponding sensing-processing workload is
\begin{equation}
    V_i^s(t)=\kappa_s D_i^s(t),
    \label{eq:sensing_workload}
\end{equation}
where \(V_i^s(t)\) is measured in CPU cycles and \(\kappa_s\) denotes the sensing-processing intensity in cycles/bit.

\subsubsection{Sensing Quality Metric}

The complex reflection coefficient is modeled using the monostatic radar link budget as
\begin{equation}
|\zeta_\ell(t)|^2
=
\frac{G_tG_r\lambda_c^2\sigma_{\ell}^{\mathrm{RCS}}(t)}
{(4\pi)^3 r_\ell^4(t)},
\label{eq:rcs_reflection}
\end{equation}
where \(G_t\) and \(G_r\) are the transmit and receive antenna gains, respectively, and \(\sigma_{\ell}^{\mathrm{RCS}}(t)\) is the radar cross section of target \(\ell\). The residual self-interference after cancellation is modeled as
\(\sigma_{\mathrm{RSI}}^2(t)=\rho_{\mathrm{SI}}p_i^{\mathrm{tx}}\), where \(\rho_{\mathrm{SI}}\) is the residual self-interference coefficient. Hence, the effective sensing SNR for target \(\ell\) at vehicle \(u_i\) is
\begin{equation}
\gamma_{i,\ell}^{s}(t)=
\frac{p_i^{\mathrm{tx}}|\zeta_\ell(t)|^2}
{\sigma_z^2+\sigma_{\mathrm{RSI}}^2(t)}.
\label{eq:sensing_snr}
\end{equation}

Following~\cite{Decarli2024JCSNRV2X}, the target-specific range and velocity CRLBs are modeled as
\begin{equation}
\epsilon_{i,\ell}^{r}(t)=
\frac{3c_0^2}
{8\pi^2M_{s,i}(k)N_{s,i}(k)\gamma_{i,\ell}^{s}(t)
\left(N_{s,i}^2(k)-1\right)\Delta f^2},
\label{eq:crlb_range}
\end{equation}
and
\begin{equation}
\epsilon_{i,\ell}^{v}(t)=
\frac{3c_0^2}
{8\pi^2f_c^2M_{s,i}(k)N_{s,i}(k)\gamma_{i,\ell}^{s}(t)
\left(M_{s,i}^2(k)-1\right)T_{\mathrm{sym}}^2}.
\label{eq:crlb_velocity}
\end{equation}
These CRLBs are used as lower-bound sensing-quality metrics evaluated from the predicted or tracked target state and the current sensing-resource allocation. They quantify the attainable estimation accuracy under the current SNR, sensing bandwidth, and coherent sensing duration, rather than assuming that delay and Doppler are perfectly known before estimation.

The detected target set is defined as
\[
\mathcal T_i(t)=
\{\ell:\gamma_{i,\ell}^{s}(t)\ge\gamma_{\mathrm{det}}^{s},
\; r_\ell(t)\le R_i^{\mathrm{sens}}\},
\]
where \(\gamma_{\mathrm{det}}^{s}\) is the sensing-detection threshold and \(R_i^{\mathrm{sens}}\) is the sensing range. The vehicle-level sensing penalty is obtained by aggregating the target-specific CRLBs:
\begin{equation}
\epsilon_i^{\mathrm{sens}}(t)=
\sum_{\ell\in\mathcal T_i(t)}
w_{i,\ell}(t)
\left(
\beta_r\epsilon_{i,\ell}^{r}(t)
+
\beta_v\epsilon_{i,\ell}^{v}(t)
\right),
\label{eq:sensing_penalty}
\end{equation}
where \(w_{i,\ell}(t)\geq0\), \(\sum_{\ell\in\mathcal T_i(t)}w_{i,\ell}(t)=1\), and larger weights are assigned to safety-critical nearby targets. In the simulations, \(w_{i,\ell}(t)\) is obtained by normalizing inverse-distance priority scores over \(\mathcal T_i(t)\).

\subsection{Communication Model}
\label{subsec:communication_model}

Each vehicle \(u_i\) generates sidelink safety packets that must be delivered to neighboring vehicles over the PC5 interface. Let \(N_i^{p}(t)\) denote the number of communication packets generated by vehicle \(u_i\) in slot \(t\), and let \(P_i^{p}\) denote the packet size in bits. The communication arrival in slot \(t\) is modeled as
\begin{equation}
A_i^{\mathrm{comm}}(t)=P_i^{p}N_i^{p}(t),
\label{eq:comm_arrival_total}
\end{equation}
where \(N_i^{p}(t)\) may represent periodic, event-driven, or mixed safety-packet generation depending on the traffic configuration. This aggregate model preserves the offered-load effect on sidelink queueing and resource allocation without requiring separate message-class scheduling.

Communication takes place over the PC5 sidelink using the SB-SPS reservation selected at control epoch \(k\). In autonomous sidelink operation, SB-SPS relies on sensing the resource pool, forming a candidate resource set, probabilistic reselection, and persistent reservation over consecutive transmissions~\cite{Todisco_ACCESS_2021}. The reselection counter and the resource keep probability are important MAC-layer parameters because they determine how long a vehicle persists on a selected resource and how frequently it re-enters the resource selection process~\cite{Zhou_TVT_2025}. For all slots \(t\in\mathcal{T}_k\), the selected communication PRB allocation is kept fixed over the corresponding reservation interval as
\begin{equation}
N_i^c(t)=N_i^c(k),
\qquad
B_i^c(t)=12\Delta f N_i^c(t),
\label{eq:comm_bandwidth}
\end{equation}
where \(B_i^c(t)\) is the sidelink communication bandwidth assigned to vehicle \(u_i\). The selected sidelink resource \(r_i(k)\), reselection counter \(RC_i(k)\), keep probability \(P_{\mathrm{keep},i}(k)\), and PRB allocation \(N_i^c(k)\) are therefore treated as epoch-level SB-SPS control variables, while the packet arrivals, channel gains, interference, and queues evolve at the slot level.

To quantify sidelink congestion, we use the channel busy ratio (CBR), defined as the fraction of resources whose received power or interference exceeds a threshold over the sensing window~\cite{Sabeeh2023CBR}. Let \(\mathcal{W}_i(t)\) denote the sensing window observed by vehicle \(u_i\). The CBR at slot \(t\) is
\begin{equation}
\mathrm{CBR}_i(t)=
\frac{
\sum_{\tau\in\mathcal{W}_i(t)}
N_{i,\mathrm{busy}}(\tau)
}{
|\mathcal{W}_i(t)|N_{\mathrm{PRB}}
},
\label{eq:comm_cbr}
\end{equation}
where \(N_{i,\mathrm{busy}}(\tau)\) is the number of PRBs sensed busy in slot \(\tau\), and \(N_{\mathrm{PRB}}\) is the total number of PRBs in the sidelink pool. A larger \(\mathrm{CBR}_i(t)\) indicates heavier channel congestion. As vehicle density increases, more PRBs are occupied, reducing the candidate resource set and increasing resource-list overlap and collision probability~\cite{Zhou_TVT_2025}.

Let \(\Omega_i(t)\) denote the intended receiver set of vehicle \(u_i\), i.e., the neighboring vehicles within the considered awareness range. For receiver \(u_j\in\Omega_i(t)\), the received sidelink SINR from transmitter \(u_i\) is modeled as
\begin{equation}
\Gamma_{i,j}^{c}(t)=
\frac{P_i^c(t)h_{i,j}(t)}
{\sigma^2+
\sum\limits_{\ell\in\mathcal{U}\setminus\{i\}}
\mathbf{1}\{r_\ell(t)=r_i(t)\}
P_\ell^c(t)h_{\ell,j}(t)},
\label{eq:comm_rx_sinr}
\end{equation}
where \(P_i^c(t)\) is the sidelink transmit power of \(u_i\), \(h_{i,j}(t)\) is the channel gain between \(u_i\) and \(u_j\), \(\sigma^2\) is the noise power, and \(r_i(t)\) denotes the SB-SPS resource used by \(u_i\) in slot \(t\). The indicator \(\mathbf{1}\{r_\ell(t)=r_i(t)\}\) captures co-channel interference from vehicles selecting the same sidelink resource.

The average received SINR over the intended receiver set is
\begin{equation}
\bar{\Gamma}_i^c(t)=
\frac{1}{|\Omega_i(t)|}
\sum_{u_j\in\Omega_i(t)}
\Gamma_{i,j}^{c}(t).
\label{eq:comm_avg_sinr}
\end{equation}
The corresponding sidelink communication rate is given by
\begin{equation}
R_i^c(t)=
B_i^c(t)\log_2\!\left(1+\bar{\Gamma}_i^c(t)\right).
\label{eq:comm_rate}
\end{equation}

For receiver \(u_j\in\Omega_i(t)\), the packet decoding indicator is defined as
\begin{equation}
z_{i,j}(t)=
\mathbf{1}\!\left\{\Gamma_{i,j}^c(t)\ge \gamma_{\mathrm{th}}^c\right\},
\label{eq:comm_success_indicator}
\end{equation}
where \(\gamma_{\mathrm{th}}^c\) is the decoding SINR threshold. The slot-wise packet reception ratio of vehicle \(u_i\) is then defined as
\begin{equation}
\mathrm{PRR}_i(t)=
\frac{1}{|\Omega_i(t)|}
\sum_{u_j\in\Omega_i(t)} z_{i,j}(t).
\label{eq:comm_prr}
\end{equation}
This definition follows the receiver-set interpretation used in V2X broadcast evaluation, where successful decoding is averaged over receivers within the effective communication range~\cite{Gu2022RLSPS}.

The effective sidelink rate and successfully delivered payload in slot \(t\) are defined as
\begin{equation}
R_i^{\mathrm{eff}}(t)=R_i^c(t)\mathrm{PRR}_i(t),
\qquad
D_i^{\mathrm{succ},c}(t)=T_{\mathrm{slot}}R_i^{\mathrm{eff}}(t).
\label{eq:comm_effective_bits}
\end{equation}
If \(D_i^{c,\min}\) is the target communication payload per slot, the minimum communication PRB demand is given by
\begin{equation}
N_i^{c,\min}(t)=
\left\lceil
\frac{D_i^{c,\min}}
{12\Delta f\,T_{\mathrm{slot}}\,\mathrm{PRR}_i(t)
\log_2(1+\bar\Gamma_i^c(t))}
\right\rceil .
\label{eq:comm_min_prb}
\end{equation}
This quantity represents the minimum sidelink PRB requirement needed to satisfy the target payload under the observed SINR and PRR conditions.

To measure the spatial coverage of reliable sidelink communication, we define the maximum reliable distance. For a given vehicle density \(\rho\), let \(\overline{\mathrm{PRR}}(d,\rho)\) denote the average packet reception ratio over transmitter--receiver pairs separated by distance \(d\). The maximum reliable sidelink distance is defined as
\begin{equation}
d_{\max}^{\mathrm{rel}}(\rho)
=
\max_{d}
\left\{
d
\; \middle| \;
\overline{\mathrm{PRR}}(d,\rho)\ge \Gamma_{\mathrm{PRR}}
\right\},
\label{eq:max_reliable_distance}
\end{equation}
where \(\Gamma_{\mathrm{PRR}}\) is the target reliability threshold. This metric is used only for performance evaluation and is not included as a Markov-game state variable.

Let \(Q_i^{\mathrm{comm}}(t)\) denote the communication queue length of vehicle \(u_i\) in bits at the beginning of slot \(t\). The queue evolves according to
\begin{equation}
Q_i^{\mathrm{comm}}(t+1)=
\left[Q_i^{\mathrm{comm}}(t)-S_i^{\mathrm{comm}}(t)\right]^+
+A_i^{\mathrm{comm}}(t),
\label{eq:comm_queue}
\end{equation}
where the effective service in slot \(t\) is
\begin{equation}
S_i^{\mathrm{comm}}(t)=
\min\!\left(Q_i^{\mathrm{comm}}(t),D_i^{\mathrm{succ},c}(t)\right).
\label{eq:comm_service}
\end{equation}
The communication delay is affected by both queueing and resource access under persistent reservation, and these components increase when resource contention becomes stronger~\cite{Zhou_TVT_2025}. In this work, the sidelink queueing delay is approximated as
\begin{equation}
T_i^{\mathrm{sl},c}(t)\approx
\frac{Q_i^{\mathrm{comm}}(t)}
{R_i^{\mathrm{eff}}(t)+\epsilon_0},
\label{eq:comm_delay}
\end{equation}
where \(\epsilon_0>0\) avoids division by zero during outage or severe congestion.

The overall communication utility is defined as
\begin{equation}
U_i^c(t)
=
\alpha_{\mathrm{prr}}\mathrm{PRR}_i(t)
+
\alpha_{\mathrm{rate}}
\frac{R_i^{\mathrm{eff}}(t)}{R_i^{\mathrm{eff},\min}},
\label{eq:comm_utility}
\end{equation}
where \(R_i^{\mathrm{eff},\min}=D_i^{c,\min}/T_{\mathrm{slot}}\), and
\(\alpha_{\mathrm{prr}},\alpha_{\mathrm{rate}}\ge0\) are weighting coefficients. The first term captures packet reception reliability, while the second term captures the achieved effective sidelink rate relative to the target rate.

\subsection{Computation Model}
\label{subsec:computation}

At slot \(t\), vehicle \(u_i\) generates a sensing workload \(V_i^s(t)\) from \eqref{eq:sensing_workload}. The communication-related processing workload is modeled as
\begin{equation}
V_i^c(t)=\kappa_c A_i^{\mathrm{comm}}(t),
\label{eq:comm_processing_workload}
\end{equation}
where \(\kappa_c\) is the required number of CPU cycles per communication bit. Computation is performed over three tiers: the on-board vehicle CPU, a MEC-enabled RSU, and the cloud, following the vehicular edge-computing architecture commonly adopted for latency-sensitive V2X services~\cite{Liu2021VECSurvey}. The RSU has a finite MEC processing capacity \(C_{\mathrm R}\) cycles/slot. Each vehicle first processes its workloads locally, with communication tasks given priority over sensing post-processing tasks. The unprocessed part is treated as overflow. Communication overflow is offloaded only to the MEC server due to its stricter latency requirement, whereas sensing overflow may be offloaded either to the MEC server or to the cloud~\cite{liu2023isac_v2x}.

\begin{algorithm}[!ht]
\caption{ISCC Offloading and Queue Update}
\label{alg:iscc_offloading_update}
\begin{algorithmic}[1]
\State \textbf{Input:} joint action $\mathbf a(k)$, slot set $\mathcal T_k$, local queues, MEC queues, $\{C_i\}_{i=1}^{N}$, $C_{\mathrm R}$
\For{each slot $t\in\mathcal T_k$}
    \For{each vehicle $u_i\in\mathcal U$}
        \State Apply $N_i^s(k)$, $N_i^c(k)$, $N_i^o(k)$, $M_i^s(k)$, $\eta_i^c(k)$, and $\eta_i^s(k)$
        \State Compute sensing data, workload, and penalty using \eqref{eq:sensing_data}--\eqref{eq:sensing_penalty}
        \State Generate communication arrivals using \eqref{eq:comm_arrival_total}
        \State Compute sidelink rate, PRR, effective rate, and delivered payload using \eqref{eq:comm_rate}--\eqref{eq:comm_effective_bits}
        \State Update the communication queue using \eqref{eq:comm_queue} and \eqref{eq:comm_service}
        \State Compute communication utility using \eqref{eq:comm_utility}
        \State Compute communication-related processing workload from the admitted communication traffic
        \State Process local communication and sensing workloads using \eqref{eq:cpu_comm_single}--\eqref{eq:cpu_sens_single}
        \State Compute communication and sensing overflow using \eqref{eq:overflow_comm_single}--\eqref{eq:overflow_sens_single}
        \State Compute V2I offloading rate, transmission delays, and transmission energies from the offloading model
    \EndFor
    \State Update MEC strict-priority service using \eqref{eq:mec_cpu_c_single} and \eqref{eq:mec_cpu_s_single}
    \State Update MEC communication and sensing queues using \eqref{eq:mec_queue_c_single} and \eqref{eq:mec_queue_s_single}
    \For{each vehicle $u_i\in\mathcal U$}
        \State Compute remote completion delays using \eqref{eq:remote_comm_delay_single} and \eqref{eq:remote_sens_delay_single}
        \State Compute communication and sensing completion delays using \eqref{eq:comp_comm_single} and \eqref{eq:comp_sens_single}
        \State Compute end-to-end communication-related delay using \eqref{eq:e2e_comm_comp_delay}
        \State Compute vehicle-side energy using \eqref{eq:total_vehicle_energy} and \eqref{eq:sensing_energy}
        \State Compute computation-side penalty using \eqref{eq:comp_penalty}
    \EndFor
\EndFor
\State \textbf{Return:} updated local queues, MEC queues, delays, energies, sensing penalties, communication utilities, and computation penalties
\end{algorithmic}
\end{algorithm}

\subsubsection{Local Processing and Overflow}

The local CPU frequency of vehicle \(u_i\) is \(F_i^{\max}\) cycles/s, and hence its per-slot computing capacity is \(C_i=F_i^{\max}T_{\mathrm{slot}}\) cycles/slot. Under communication-priority processing, the locally served communication and sensing workloads are
\begin{align}
f_i^c(t)
&=
\min\!\left\{V_i^c(t),\, C_i\right\},
\label{eq:cpu_comm_single}
\\
f_i^s(t)
&=
\min\!\left\{V_i^s(t),\, [C_i-f_i^c(t)]^+\right\},
\label{eq:cpu_sens_single}
\end{align}
where \([x]^+=\max\{x,0\}\). Therefore, the total locally processed workload is \(V_i^{\mathrm{loc}}(t)=f_i^c(t)+f_i^s(t)\). The corresponding local completion times are \(T_i^{\mathrm{loc},c}(t)=f_i^c(t)/F_i^{\max}\) and \(T_i^{\mathrm{loc},s}(t)=(f_i^c(t)+f_i^s(t))/F_i^{\max}\), where the latter includes the waiting effect induced by communication-priority processing.

Using the standard DVFS-based CPU energy model~\cite{Mao2017MEC}, the local computation energy is given by
\begin{equation}
E_i^{\mathrm{loc}}(t)
=
\kappa_i V_i^{\mathrm{loc}}(t)
\left(f_i^{\mathrm{eff}}(t)\right)^2,
\qquad
f_i^{\mathrm{eff}}(t)=\frac{V_i^{\mathrm{loc}}(t)}{T_{\mathrm{slot}}},
\label{eq:local_energy_single}
\end{equation}
where \(\kappa_i\) is the effective switched-capacitance coefficient and \(f_i^{\mathrm{eff}}(t)\le F_i^{\max}\). Equivalently, \eqref{eq:local_energy_single} can be written as
\(E_i^{\mathrm{loc}}(t)=\kappa_i\big(V_i^{\mathrm{loc}}(t)/T_{\mathrm{slot}}\big)^3T_{\mathrm{slot}}\). The resulting communication and sensing overflow workloads are
\begin{align}
V_i^{\mathrm{ov},c}(t)
&=
\big[V_i^c(t)-f_i^c(t)\big]^+,
\label{eq:overflow_comm_single}
\\
V_i^{\mathrm{ov},s}(t)
&=
\big[V_i^s(t)-f_i^s(t)\big]^+.
\label{eq:overflow_sens_single}
\end{align}
This overflow-triggered formulation follows the partial-offloading principle in vehicular MEC while restricting the remote destination according to the ISCC task class~\cite{Liu2022UAVISAC}.

\subsubsection{Offloading Model}

At control epoch \(k\), the communication offloading decision is denoted by \(\eta_i^c(k)\in\{0,1\}\), where \(\eta_i^c(k)=1\) indicates MEC offloading. The sensing offloading decision is denoted by \(\eta_i^s(k)\in\{0,1,2\}\), corresponding to no offloading, MEC offloading, and cloud offloading, respectively. These epoch-level decisions remain fixed for all slots \(t\in\mathcal T_k\). Communication overflow is eligible for MEC offloading only when \(V_i^{\mathrm{ov},c}(t)>0\), while sensing overflow is eligible for MEC or cloud offloading only when \(V_i^{\mathrm{ov},s}(t)>0\). When the MEC backlog is high, the policy can suppress additional communication offloading through \(\eta_i^c(k)=0\) and route sensing overflow to the cloud through \(\eta_i^s(k)=2\), thereby reducing further MEC congestion.

The overflow workloads are mapped into class-wise uplink payloads as \(D_i^{\mathrm{off},c}(t)=\xi_cV_i^{\mathrm{ov},c}(t)\) and \(D_i^{\mathrm{off},s}(t)=\xi_sV_i^{\mathrm{ov},s}(t)\), where \(\xi_c\) and \(\xi_s\) are bits/cycle conversion factors. Since \(N_i^o(k)\) is the V2I offloading PRB allocation selected at epoch \(k\), the offloading bandwidth is \(B_i^{\mathrm{off}}(t)=12\Delta f N_i^o(k)\) for \(t\in\mathcal T_k\). For the set of offloading vehicles \(\mathcal{U}^{\mathrm{off}}(t)\), the uplink SINR from \(u_i\) to the RSU is
\begin{equation}
\mathrm{SINR}_{i}^{\mathrm{off}}(t)
=
\frac{
p_i^{\mathrm{off}}g_{i}^{\mathrm{RSU}}(t)
}{
\sum\limits_{m\in \mathcal{U}^{\mathrm{off}}(t)\setminus\{i\}}
p_m^{\mathrm{off}}g_{m}^{\mathrm{RSU}}(t)
+
N_0 B_i^{\mathrm{off}}(t)
},
\label{eq:sinr_off_single}
\end{equation}
where \(p_i^{\mathrm{off}}\) is the uplink offloading power, \(g_i^{\mathrm{RSU}}(t)\) is the uplink channel gain, and \(N_0\) is the noise power spectral density. The corresponding offloading rate is
\(R_i^{\mathrm{off}}(t)=B_i^{\mathrm{off}}(t)\log_2(1+\mathrm{SINR}_i^{\mathrm{off}}(t))\)~\cite{Tan2022OFDMAEdge}. The class-wise uplink transmission delays are
\(T_i^{\mathrm{tx},c}(t)=D_i^{\mathrm{off},c}(t)/R_i^{\mathrm{off}}(t)\) and
\(T_i^{\mathrm{tx},s}(t)=D_i^{\mathrm{off},s}(t)/R_i^{\mathrm{off}}(t)\), with transmission energies
\(E_i^{\mathrm{tx},c}(t)=p_i^{\mathrm{off}}T_i^{\mathrm{tx},c}(t)\) and
\(E_i^{\mathrm{tx},s}(t)=p_i^{\mathrm{off}}T_i^{\mathrm{tx},s}(t)\).

\subsubsection{MEC and Cloud Processing}

The RSU maintains two CPU-cycle queues, \(L_{\mathrm R}^c(t)\) and \(L_{\mathrm R}^s(t)\), for communication and sensing overflow. Since communication tasks are more latency-sensitive, the MEC server applies strict-priority service:
\begin{align}
C_{\mathrm R}^c(t)
&=
\min\!\left\{L_{\mathrm R}^c(t),\, C_{\mathrm R}\right\},
\label{eq:mec_cpu_c_single}
\\
C_{\mathrm R}^s(t)
&=
\min\!\left\{L_{\mathrm R}^s(t),\, [C_{\mathrm R}-C_{\mathrm R}^c(t)]^+\right\}.
\label{eq:mec_cpu_s_single}
\end{align}
The MEC queue dynamics are given by
\begin{align}
L_{\mathrm R}^c(t+1)
&=
\big[L_{\mathrm R}^c(t)-C_{\mathrm R}^c(t)\big]^+
+
\sum_{i=1}^{N}
\mathbf{1}_{\{\eta_i^c(k)=1\}}V_i^{\mathrm{ov},c}(t),
\label{eq:mec_queue_c_single}
\\
L_{\mathrm R}^s(t+1)
&=
\big[L_{\mathrm R}^s(t)-C_{\mathrm R}^s(t)\big]^+
+
\sum_{i=1}^{N}
\mathbf{1}_{\{\eta_i^s(k)=1\}}V_i^{\mathrm{ov},s}(t),
\label{eq:mec_queue_s_single}
\end{align}
for \(t\in\mathcal T_k\). The MEC queueing delay is estimated using a backlog-to-service approximation as
\(T_{\mathrm R}^{\mathrm{que},c}(t)\approx L_{\mathrm R}^c(t)T_{\mathrm{slot}}/C_{\mathrm R}\) and
\(T_{\mathrm R}^{\mathrm{que},s}(t)\approx (L_{\mathrm R}^c(t)+L_{\mathrm R}^s(t))T_{\mathrm{slot}}/C_{\mathrm R}\), where sensing experiences the accumulated effect of the higher-priority communication backlog~\cite{liu2023isac_v2x}. This approximation is used for delay estimation and does not constitute a formal queue-stability result.

For class \(q\in\{c,s\}\), the MEC execution delay is approximated as
\(T_i^{\mathrm{mec},q}(t)\approx V_i^{\mathrm{ov},q}(t)T_{\mathrm{slot}}/C_{\mathrm R}\), and the total MEC-side delay is
\(T_i^{\mathrm{edge},q}(t)=T_{\mathrm R}^{\mathrm{que},q}(t)+T_i^{\mathrm{mec},q}(t)\). Only sensing overflow can be sent to the cloud. When \(\eta_i^s(k)=2\), the sensing overflow is uploaded to the RSU and forwarded over the backhaul. Cloud-offloaded sensing tasks use the RSU only as a forwarding node and are not inserted into the MEC computing queue. The cloud-side delay is modeled by \(T^{\mathrm{bh}}+T^{\mathrm{cl}}\), where \(T^{\mathrm{bh}}\) is the backhaul delay and \(T^{\mathrm{cl}}\) is the cloud execution delay.

\subsubsection{Completion Delay, Energy, and Computation Cost}

The remote completion delays of communication and sensing overflow are given by
\begin{align}
T_i^{\mathrm{rem},c}(t)
&=
\mathbf{1}_{\{\eta_i^c(k)=1\}}
\left(T_i^{\mathrm{tx},c}(t)+T_i^{\mathrm{edge},c}(t)\right),
\label{eq:remote_comm_delay_single}
\\
T_i^{\mathrm{rem},s}(t)
&=
\mathbf{1}_{\{\eta_i^s(k)=1\}}
\left(T_i^{\mathrm{tx},s}(t)+T_i^{\mathrm{edge},s}(t)\right)
\nonumber\\
&\quad+
\mathbf{1}_{\{\eta_i^s(k)=2\}}
\left(T_i^{\mathrm{tx},s}(t)+T^{\mathrm{bh}}+T^{\mathrm{cl}}\right).
\label{eq:remote_sens_delay_single}
\end{align}
Since only the workload exceeding the local CPU service capacity is offloaded, local processing and remote overflow processing may proceed over different execution paths. Hence, the completion time is determined by the slower branch, as in split-processing offloading models~\cite{Liu2022UAVISAC}:
\begin{align}
T_i^{\mathrm{comp},c}(t)
&=
\max\!\left\{
T_i^{\mathrm{loc},c}(t),
T_i^{\mathrm{rem},c}(t)
\right\},
\label{eq:comp_comm_single}
\\
T_i^{\mathrm{comp},s}(t)
&=
\max\!\left\{
T_i^{\mathrm{loc},s}(t),
T_i^{\mathrm{rem},s}(t)
\right\}.
\label{eq:comp_sens_single}
\end{align}
The end-to-end communication-related delay is
\begin{equation}
T_i^{c,\mathrm{tot}}(t)
=
T_i^{\mathrm{sl},c}(t)
+
T_i^{\mathrm{comp},c}(t),
\label{eq:e2e_comm_comp_delay}
\end{equation}
where \(T_i^{\mathrm{sl},c}(t)\) is the sidelink communication delay defined in Section~\ref{subsec:communication_model}.

The total vehicle-side energy consumption in slot \(t\) is
\begin{equation}
E_i^{\mathrm{tot}}(t)
=
E_i^{\mathrm{loc}}(t)
+
E_i^{\mathrm{tx},c}(t)
+
E_i^{\mathrm{tx},s}(t)
+
E_i^{\mathrm{sens}}(t),
\label{eq:total_vehicle_energy}
\end{equation}
where \(E_i^{\mathrm{loc}}(t)\) is the local computation energy, \(E_i^{\mathrm{tx},c}(t)\) and \(E_i^{\mathrm{tx},s}(t)\) are the transmission energies for communication- and sensing-related offloading, respectively, and \(E_i^{\mathrm{sens}}(t)\) is the sensing transmission energy. For \(t\in\mathcal{T}_k\),
\begin{equation}
E_i^{\mathrm{sens}}(t)
=
p_i^{\mathrm{sens}}M_i^s(k)T_{\mathrm{sym}},
\label{eq:sensing_energy}
\end{equation}
where \(p_i^{\mathrm{sens}}\), \(M_i^s(k)\), and \(T_{\mathrm{sym}}\) denote the sensing transmit power, the number of sensing OFDM symbols selected at control epoch \(k\), and the OFDM symbol duration, respectively.

Finally, the computation-side penalty is defined as
\begin{equation}
\psi_i^{\mathrm{comp}}(t)
=
\alpha_{\mathrm{dc}}
\frac{T_i^{\mathrm{comp},c}(t)}{\delta_i^c}
+
\alpha_{\mathrm{ds}}
\frac{T_i^{\mathrm{comp},s}(t)}{\delta_i^s}
+
\alpha_{\mathrm{e}}
\frac{E_i^{\mathrm{tot}}(t)}{E_i^{\max}},
\label{eq:comp_penalty}
\end{equation}
where \(\alpha_{\mathrm{dc}},\alpha_{\mathrm{ds}},\alpha_{\mathrm{e}}\ge0\) weight communication-computation delay, sensing completion delay, and vehicle-side energy consumption, respectively.

\section{Problem Formulation}
\label{sec:problem}

Using the models in Section~\ref{sec:system_model}, we seek a cross-layer policy that jointly selects the SB-SPS reservation update, PC5 sensing--communication split, V2I offloading allocation, and overflow-handling decision. The objective is to balance sensing accuracy, sidelink utility, computation delay, and vehicle-side energy. Since these terms have different physical units, we aggregate only their normalized dimensionless forms. Let
\(\widetilde{\epsilon}_i^{\mathrm{sens}}(t)\),
\(\widetilde{\Phi}_i^c(t)\), and
\(\widetilde{\psi}_i^{\mathrm{comp}}(t)\) denote the normalized sensing penalty, communication deficiency, and computation-side penalty obtained from \eqref{eq:sensing_penalty}, \eqref{eq:comm_utility}, and \eqref{eq:comp_penalty}, respectively. These quantities are clipped to \([0,1]\), where \(\widetilde{\Phi}_i^c(t)\) increases when the achieved communication utility falls below its target level. The slot-wise cross-layer cost of vehicle \(u_i\) is defined as
\begin{equation}
\ell_i(t)
=
\omega_s\widetilde{\epsilon}_i^{\mathrm{sens}}(t)
+
\omega_c\widetilde{\Phi}_i^c(t)
+
\omega_{\mathrm{comp}}\widetilde{\psi}_i^{\mathrm{comp}}(t),
\label{eq:inst_cost}
\end{equation}
where \(\omega_s,\omega_c,\omega_{\mathrm{comp}}\ge0\) and
\(\omega_s+\omega_c+\omega_{\mathrm{comp}}=1\). Thus, \eqref{eq:inst_cost} combines only bounded dimensionless costs rather than raw CRLB, rate, delay, or energy terms.

The long-term control problem is formulated as
\begin{equation}
\label{eq:P0}
\mathbf{P0:}\qquad
\min_{\Pi}
\limsup_{T\to\infty}
\frac{1}{NT}
\sum_{t=1}^{T}\sum_{i=1}^{N}
\mathbb{E}\!\left[\ell_i(t)\right],
\end{equation}
subject to
\begin{equation*}
\begin{aligned}
\textbf{C1:}\quad
&N_i^s(k)+N_i^c(k)\le N_i^{\mathrm{sl}}(k), 
&&\forall i,k,\\
\textbf{C2:}\quad
&0\le N_i^o(k)\le N_i^{o,\max}(k),
&&\forall i,k,\\
\textbf{C3:}\quad
&r_i(k)\in\mathcal R_i^{\mathrm{cand}}(k),
&&\forall i,k,\\
\textbf{C4:}\quad
&RC_i(k)\in\mathcal C_{RC},\quad 
P_{\mathrm{keep},i}(k)\in\mathcal P_{\mathrm{keep}},
&&\forall i,k,\\
\textbf{C5:}\quad
&1\le M_i^s(k)\le 14,
&&\forall i,k,\\
\textbf{C6:}\quad
&\eta_i^c(k)\in\{0,1\},\quad
\eta_i^s(k)\in\{0,1,2\},
&&\forall i,k,\\
\textbf{C7:}\quad
&T_i^{c,\mathrm{tot}}(t)\le \delta_i^c,
&&\forall i,t,\\
\textbf{C8:}\quad
&T_i^{\mathrm{comp},s}(t)\le \delta_i^s,
&&\forall i,t,\\
\textbf{C9:}\quad
&E_i^{\mathrm{tot}}(t)\le E_i^{\max},
&&\forall i,t.
\end{aligned}
\end{equation*}

Here, \(\textbf{C1}\) separates PC5 sensing and communication resources, \(\textbf{C2}\) limits V2I offloading resources, \(\textbf{C3}\)--\(\textbf{C6}\) define feasible SB-SPS and overflow-handling actions, and \(\textbf{C7}\)--\(\textbf{C9}\) impose communication-delay, sensing-completion-delay, and energy constraints. The offloading decisions are influenced by MEC backlog and service capacity through the queueing-delay and computation-cost terms, thereby discouraging persistent MEC congestion. We therefore use \(\textbf{P0}\) as the analytical objective and model the control problem as a cooperative Markov game. MARL is suitable for this setting because it supports distributed decisions under local observations, mobility, and coupled resource interactions~\cite{Althamary_WCMC_2019}, and has been used for SPS-based C-V2X resource reselection~\cite{Gu2022RLSPS}.

\begin{figure*}[!t]
\centering

\begin{minipage}[t]{0.64\textwidth}
    \centering
    \vspace{0pt}
    \includegraphics[width=11.5cm,height=6.9cm]{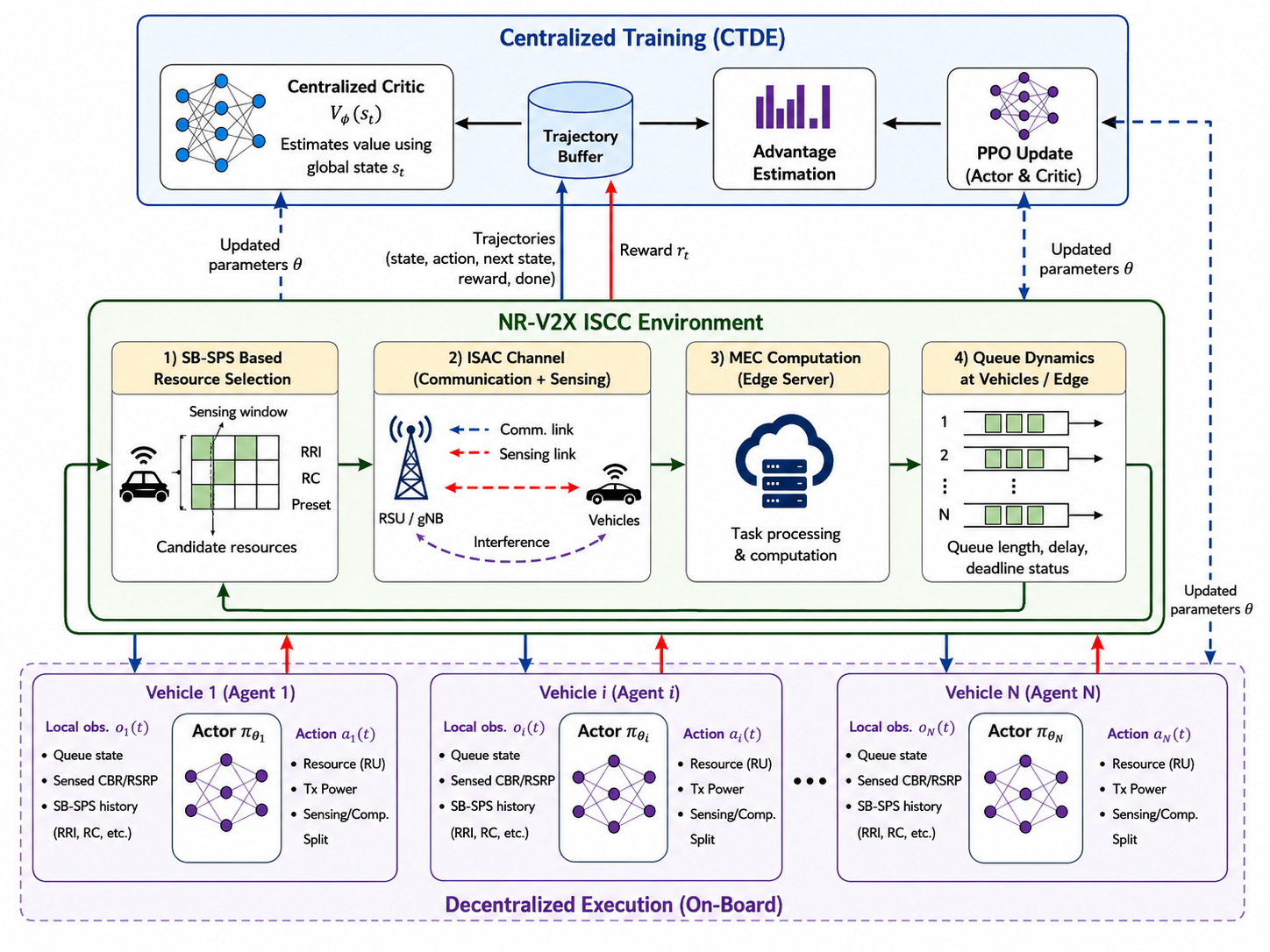}
    \caption{Proposed MAPPO-based ISCC architecture.}
    \label{fig:mappo_architecture}
\end{minipage}
\hfill
\begin{minipage}[t]{0.31\textwidth}
    \centering
    \vspace{1pt}
    \captionof{table}{Simulation Parameters}
    \label{tab:iscc_params}
    \renewcommand{\arraystretch}{0.88}
    \setlength{\tabcolsep}{1.8pt}
    {\fontsize{7.5}{7.5}\selectfont
    \begin{tabular}{p{0.52\linewidth}p{0.36\linewidth}}
    \hline
    \textbf{Parameter} & \textbf{Value} \\
    \hline
    Carrier frequency, $f_c$ & 5.9 GHz \\
    System bandwidth, $B$ & 10 MHz \\
    Modulation/coding & MCS 9 \\
    Packet size, $Z$ & 190 bytes \\
    Message rate & 10, 20, 50 Hz \\
    Tx power, $p_i^{\mathrm{tx}}$ & 23 dBm \\
    RSRP threshold & $-128$ dBm \\
    SNR threshold & 8 dB \\
    $\rho_{\mathrm{SI}}$, RCS & $10^{-7}$, 10 dBsm \\
    Antenna gain & 3 dB \\
    Max. speed & 70 km/h \\
    Veh. density, $\rho$ & 20--100 veh/km \\
    $T_{\mathrm{sen}}$ & 1000 ms \\
    $T_{\mathrm{sel}}$ & 100 ms \\
    $\mathcal C_{RC}(T_{\mathrm{RRI}})$ & RRI-dependent \\
    Vehicle CPU, $F_i^{\max}$ & 2 GHz \\
    $(\delta_i^c,\delta_i^s)$ & $(20,50)$ ms \\
    $(\omega_s,\omega_c,\omega_{\mathrm{comp}})$ & $(0.30,0.35,0.35)$ \\
$(\alpha_{\mathrm{prr}},\alpha_{\mathrm{rate}})$ & $(0.5,0.5)$ \\
    Episodes, $N_{\mathrm{epi}}$ & 500 \\
    Steps/episode & 40 \\
    Actor/critic LR & $(10^{-3},10^{-3})$ \\
    $\gamma,\lambda,\epsilon_{\mathrm{clip}}$ & $0.99,0.95,0.2$ \\
    \hline
    \end{tabular}
    }
\end{minipage}

\end{figure*}

\section{Proposed MAPPO-SPS Framework}
\label{sec:markov_game}

\subsection{Cooperative Markov Game Formulation}
\label{subsec:markov_game_formulation}

Problem~\textbf{P0} involves coupled slot-level sensing, communication, and computation dynamics. In NR-V2X Mode~2, each vehicle makes resource decisions using local sensing information, while its action affects neighboring vehicles through interference, packet reception, CBR, and MEC workload. We therefore model the ISCC-aware scheduling problem as a cooperative partially observable Markov game, equivalently a multi-agent POMDP with a shared reward. The problem is solved using MAPPO under the centralized-training decentralized-execution (CTDE) paradigm~\cite{Yu2022MAPPO}, where decentralized actors use local observations during execution and the centralized critic is used only during training. This formulation is consistent with SPS-based V2X MARL resource selection~\cite{Gu2022RLSPS}.

Let \(\mathcal{U}=\{u_1,\ldots,u_N\}\) denote the set of vehicular agents. The Markov game is defined as
\begin{equation}
\mathcal{G}=
\left(
\mathcal{U},
\mathcal{S},
\{\mathcal{O}_i\}_{i=1}^{N},
\{\mathcal{A}_i\}_{i=1}^{N},
\mathcal{P},
r,
\gamma
\right),
\label{eq:markov_game_tuple}
\end{equation}
where \(\mathcal{S}\) is the global state space, \(\mathcal{O}_i\) and \(\mathcal{A}_i\) are the local observation and action spaces of vehicle \(u_i\), \(\mathcal{P}\) is the transition kernel, \(r\) is the cooperative reward, and \(\gamma\in(0,1)\) is the discount factor. The decision index \(k\) denotes an SB-SPS control epoch, while slot-level ISCC dynamics evolve over \(t\in\mathcal{T}_k\). The CTDE architecture is shown in Fig.~\ref{fig:mappo_architecture}, where decentralized vehicle actors execute actions using local observations and the centralized critic evaluates the joint behavior only during training.

During centralized training, the critic observes the global state
\begin{equation}
s(k)=
\left(
\{o_i(k)\}_{i=1}^{N},
L_{\mathrm R}^{c}(k),
L_{\mathrm R}^{s}(k)
\right),
\label{eq:marl_state}
\end{equation}
where \(L_{\mathrm R}^{c}(k)\) and \(L_{\mathrm R}^{s}(k)\) are the RSU-side MEC backlogs for communication- and sensing-related workloads. The local observation of vehicle \(u_i\) is
\begin{equation}
\begin{aligned}
o_i(k)=\Big(
&Q_i^{\mathrm{comm}}(k),
V_i^s(k),
V_i^{\mathrm{ov},c}(k),
V_i^{\mathrm{ov},s}(k),\\
&\mathrm{CBR}_i(k),
\mathrm{PRR}_i(k),
R_i^{\mathrm{eff}}(k),
\bar{\Gamma}_i^c(k),\\
&\bar{RC}_i(k),
g_i^{\mathrm{RSU}}(k),
b_{\mathrm R}^{c}(k),
b_{\mathrm R}^{s}(k)
\Big).
\end{aligned}
\label{eq:marl_obs}
\end{equation}
Here, \(Q_i^{\mathrm{comm}}(k)\), \(V_i^s(k)\), \(V_i^{\mathrm{ov},c}(k)\), and \(V_i^{\mathrm{ov},s}(k)\) describe the local queue and workload state; \(\mathrm{CBR}_i(k)\), \(\mathrm{PRR}_i(k)\), \(R_i^{\mathrm{eff}}(k)\), and \(\bar{\Gamma}_i^c(k)\) summarize the observed sidelink condition; \(\bar{RC}_i(k)\) denotes the remaining reselection-counter state of the current SB-SPS reservation; \(g_i^{\mathrm{RSU}}(k)\) is the observed V2I channel gain; and \(b_{\mathrm R}^{c}(k)\), \(b_{\mathrm R}^{s}(k)\) are limited RSU-side backlog indicators. The candidate resource set \(\mathcal R_i^{\mathrm{cand}}(k)\), obtained from the Mode~2 sensing and selection procedure, is used as an action mask rather than as a state variable.
At each control epoch \(k\), vehicle \(u_i\) selects the extended SB-SPS-compatible ISCC action
\begin{equation}
\begin{aligned}
a_i(k)=\big(&r_i(k),RC_i(k),P_{\mathrm{keep},i}(k),
N_i^s(k),N_i^c(k),\\
&N_i^o(k),M_i^s(k),\eta_i^c(k),\eta_i^s(k)\big).
\end{aligned}
\label{eq:marl_action}
\end{equation}
The feasible set is defined as
\begin{equation}
\begin{aligned}
\mathcal A_i(k)=\{a_i(k)\mid\;&
r_i(k)\in\mathcal R_i^{\mathrm{cand}}(k),\;
RC_i(k)\in\mathcal C_{RC}(T_{\mathrm{RRI}}),\\
&\hspace{-2em}
P_{\mathrm{keep},i}(k)\in\mathcal P_{\mathrm{keep}},
\;N_i^s(k)+N_i^c(k)\le N_i^{\mathrm{sl}}(k),\\
&\hspace{-2em}
0\le N_i^o(k)\le N_i^{o,\max}(k),
\;M_i^s(k)\in\{1,\ldots,14\},\\
&\hspace{-2em}
\eta_i^c(k)\in\{0,1\},
\;\eta_i^s(k)\in\{0,1,2\}\}.
\end{aligned}
\label{eq:feasible_action_set}
\end{equation}
Here, \(\mathcal R_i^{\mathrm{cand}}(k)\) is obtained from the Mode~2 sensing and selection procedure and used as an action mask. The PC5 variables \(r_i(k)\), \(RC_i(k)\), \(P_{\mathrm{keep},i}(k)\), \(N_i^s(k)\), \(N_i^c(k)\), and \(M_i^s(k)\) control the extended sidelink reservation and sensing--communication partition, whereas \(N_i^o(k)\), \(\eta_i^c(k)\), and \(\eta_i^s(k)\) control V2I offloading and overflow routing. The action is updated only when SB-SPS reselection or re-evaluation is triggered; otherwise, the previous reservation is retained over \(\mathcal T_k\).
After the joint action \(\mathbf a(k)=\{a_i(k)\}_{i=1}^{N}\) is executed, the system evolves according to the transition kernel \(s(k+1)\sim\mathcal{P}(s(k+1)\mid s(k),\mathbf a(k))\).
The transition is induced by the analytical models in Section~\ref{sec:system_model}. To align learning with \textbf{P0}, the cooperative reward is defined as the negative average normalized slot-level cost over the control interval:
\begin{equation}
r(k)=
-\frac{1}{N|\mathcal{T}_k|}
\sum_{i=1}^{N}
\sum_{t\in\mathcal{T}_k}
\ell_i(t),
\label{eq:marl_reward}
\end{equation}
where \(\ell_i(t)\) is defined in \eqref{eq:inst_cost}. Maximizing the expected return therefore minimizes the long-term ISCC cost in \textbf{P0}. Each vehicle follows a decentralized policy \(\pi_i(a_i(k)\mid o_i(k))\), and the joint policy is factorized as
\begin{equation}
\boldsymbol{\pi}(\mathbf a(k)\mid\mathbf o(k))
=
\prod_{i=1}^{N}
\pi_i(a_i(k)\mid o_i(k)).
\label{eq:joint_policy}
\end{equation}
The cooperative learning objective is
\begin{equation}
\max_{\boldsymbol{\pi}}
\;
\mathbb{E}_{\boldsymbol{\pi}}
\left[
\sum_{k=0}^{\infty}
\gamma^k r(k)
\right].
\label{eq:marl_objective}
\end{equation}

\subsection{MAPPO-Based Training}
\label{subsec:mappo_training}

The cooperative Markov game is solved using MAPPO under CTDE~\cite{Yu2022MAPPO}. Let \(\theta_i\) denote the actor parameters of vehicle \(u_i\), and let \(\phi\) denote the centralized critic parameters. For vehicle \(u_i\), the PPO probability ratio is
\begin{equation}
\rho_i(k)=
\frac{
\pi_{\theta_i}\!\left(a_i(k)\mid o_i(k)\right)
}{
\pi_{\theta_i^{\mathrm{old}}}\!\left(a_i(k)\mid o_i(k)\right)
}.
\label{eq:ppo_ratio}
\end{equation}
The clipped actor objective follows the PPO surrogate formulation~\cite{Schulman2017PPO}:
\begin{equation}
\small
\begin{aligned}
\mathcal{L}_{i}^{\mathrm{clip}}(\theta_i)
=
\mathbb{E}\!\Big[
\min\!\big(
&\rho_i(k)\hat A(k),\\
&\operatorname{clip}(\rho_i(k),1-\epsilon,1+\epsilon)\hat A(k)
\big)
\Big],
\end{aligned}
\label{eq:ppo_clip}
\end{equation}
where \(\epsilon\) is the clipping parameter and \(\hat A(k)\) is the advantage estimate. The centralized critic estimates \(V_{\phi}(s(k))\), and the temporal-difference residual is
\begin{equation}
\delta(k)=
r(k)+\gamma V_{\phi}(s(k+1))-V_{\phi}(s(k)).
\label{eq:td_error}
\end{equation}
The generalized advantage estimate is
\begin{equation}
\hat A(k)=
\sum_{\ell=0}^{\infty}
(\gamma\lambda)^{\ell}\delta(k+\ell),
\label{eq:gae}
\end{equation}
where \(\lambda\in[0,1]\) is the GAE parameter. Since the reward is shared, the same team-level advantage is used for all decentralized actors.

The critic is trained by minimizing
\begin{equation}
\mathcal{L}^{V}(\phi)
=
\mathbb{E}
\left[
\left(
V_{\phi}(s(k))-\hat V(k)
\right)^2
\right],
\label{eq:value_loss}
\end{equation}
where \(\hat V(k)\) is the target return. The MAPPO objective is
\begin{equation}
\mathcal{L}^{\mathrm{MAPPO}}
=
\sum_{i=1}^{N}
\mathcal{L}_{i}^{\mathrm{clip}}(\theta_i)
-
c_v\mathcal{L}^{V}(\phi)
+
c_e\mathcal{H}(\pi_{\theta}),
\label{eq:mappo_obj}
\end{equation}
where \(c_v\) and \(c_e\) are the value-loss and entropy coefficients. The training and execution procedure is summarized in Algorithm~\ref{alg:mappo_iscc_sps}.

\begin{algorithm}[!ht]
\caption{Proposed MAPPO-SPS Algorithm}
\label{alg:mappo_iscc_sps}
\begin{algorithmic}[1]
\State \textbf{Input:} $\mathcal U$, admissible SB-SPS sets, training parameters $\gamma,\lambda,\epsilon_{\mathrm{clip}}$
\State Initialize decentralized actors $\{\pi_{\theta_i}\}_{i=1}^{N}$ and centralized critic $V_{\phi}$
\For{each episode}
    \State Initialize mobility, channels, queues, MEC backlog, and SB-SPS reservations
    \For{each control epoch $k$}
        \For{each vehicle $u_i$}
            \State Observe $o_i(k)$ using \eqref{eq:marl_obs}
            \State Obtain $\mathcal R_i^{\mathrm{cand}}(k)$ from the SB-SPS sensing/selection window
            \State Construct the masked action domain using $\mathcal R_i^{\mathrm{cand}}(k)$, $\mathcal C_{RC}(T_{\mathrm{RRI}})$, $\mathcal P_{\mathrm{keep}}$, and the PRB/offloading constraints
            \If{$\bar{RC}_i(k)=0$ or re-evaluation is triggered}
                \State Sample $a_i(k)\sim\pi_{\theta_i}^{\mathrm{old}}(\cdot|o_i(k))$ with infeasible actions masked
            \Else
                \State Retain the previous SB-SPS reservation and update only admissible cross-layer components
            \EndIf
        \EndFor
        \State Execute $\mathbf a(k)$ and simulate slot-level ISCC dynamics over $\mathcal T_k$
        \State Compute $r(k)$ using \eqref{eq:marl_reward} and construct $s(k+1)$
        \State Store transition $(s(k),\mathbf o(k),\mathbf a(k),r(k),s(k+1),\mathbf o(k+1))$
    \EndFor
    \State Compute advantages using \eqref{eq:td_error}--\eqref{eq:gae}
    \State Update actors using \eqref{eq:ppo_ratio}--\eqref{eq:ppo_clip}
    \State Update critic using \eqref{eq:value_loss}
    \State Set $\theta_i^{\mathrm{old}}\leftarrow\theta_i$, $\forall i$
\EndFor
\State \textbf{Output:} trained decentralized policies $\{\pi_{\theta_i}\}_{i=1}^{N}$
\end{algorithmic}
\end{algorithm}

\subsection{Benchmarking Algorithms}
\label{subsec:benchmarks}

The proposed MAPPO-SPS framework is compared with three baselines under identical network topology, channel, traffic, and SB-SPS operating conditions. The two greedy baselines are non-learning heuristics designed to isolate sensing-centric and computation-centric behavior, while MA-A2C-SPS provides the learning-based baseline under the same observation, action, and reward definitions.

\subsubsection{\textbf{SCG-SPS: Sensing-Centric Greedy SB-SPS}}

follows the same Mode~2 candidate-resource selection and reservation-persistence rules as MAPPO-SPS, but greedily minimizes the sensing cost in \eqref{eq:sensing_penalty}. At control epoch \(k\), vehicle \(u_i\) selects
\begin{equation}
a_i^{\mathrm{SCG}}(k)
=
\arg\min_{a_i(k)\in\mathcal A_i(k)}
\sum_{t\in\mathcal T_k}
\epsilon_i^{\mathrm{sens}}(t),
\label{eq:scg_sbsps_policy}
\end{equation}
where \(\mathcal A_i(k)\) is the feasible extended SB-SPS action set. 
\subsubsection{\textbf{CCG-SPS: Computing-Centric Greedy SB-SPS}}

CCG-SPS uses the same extended SB-SPS reservation and persistence structure, but prioritizes computation-side performance. At each control epoch, vehicle \(u_i\) chooses the feasible action minimizing the accumulated computation cost:
\begin{equation}
a_i^{\mathrm{CG}}(k)
=
\arg\min_{a_i(k)\in\mathcal A_i(k)}
\sum_{t\in\mathcal T_k}
\psi_i^{\mathrm{comp}}(t),
\label{eq:cg_sbsps_policy}
\end{equation}
where \(\psi_i^{\mathrm{comp}}(t)\) is defined in \eqref{eq:comp_penalty}.
\subsubsection{\textbf{MA-A2C-SPS: Multi-Agent Advantage Actor--Critic}}
MA-A2C-SPS is a learning-based baseline that uses the same
observation space, action space, and cooperative reward in
(\ref{eq:marl_reward}) as MAPPO-SPS. Thus, it optimizes the
same cross-layer objective derived from Problem~\(\mathbf{P0}\),
but updates the policy using a standard multi-agent advantage
actor--critic rule instead of the clipped PPO objective.

\subsection{Complexity Analysis}
\label{subsec:complexity}

The complexity of MAPPO-SPS is dominated by actor--critic updates during training, while online execution only requires decentralized actor inference~\cite{li2025mappo_mec}. Let the actor and critic have \(L_a\) and \(L_c\) fully connected layers, with \(\beta_{\ell}^{a}\) and \(\beta_{\ell}^{c}\) neurons in layer \(\ell\), respectively. The actor and critic forward-pass complexities are
\[
C_a=\mathcal{O}\!\left(\sum_{\ell=1}^{L_a}\beta_{\ell-1}^{a}\beta_{\ell}^{a}\right),
\qquad
C_c=\mathcal{O}\!\left(\sum_{\ell=1}^{L_c}\beta_{\ell-1}^{c}\beta_{\ell}^{c}\right).
\]
The backward pass has the same asymptotic order up to a constant factor. For \(N_{\mathrm{epi}}\) episodes, \(N_{\mathrm{upd}}\) PPO update epochs, and minibatch size \(N_{\mathrm{batch}}\), the training complexity is
\(\mathcal{O}\!\left(N_{\mathrm{epi}}N_{\mathrm{upd}}N_{\mathrm{batch}}(NC_a+C_c)\right)\), where \(NC_a\) accounts for the actor updates of \(N\) vehicles and \(C_c\) accounts for the centralized critic update. Scalability is mainly affected during centralized training because the critic input grows with the number of vehicles. During online execution, the critic is discarded and each vehicle evaluates only its local actor; hence, the deployed decision step is a lightweight neural-network forward pass and does not require centralized optimization or iterative search within an SB-SPS control epoch.

\section{Simulation Results}
\label{sec:results}

\begin{figure*}[!t]
\centering
\setlength{\abovecaptionskip}{2pt}
\setlength{\belowcaptionskip}{-5pt}
\captionsetup[subfloat]{font=scriptsize,labelfont=scriptsize}

\subfloat[Average PRR.]{
    \includegraphics[width=0.34\textwidth]{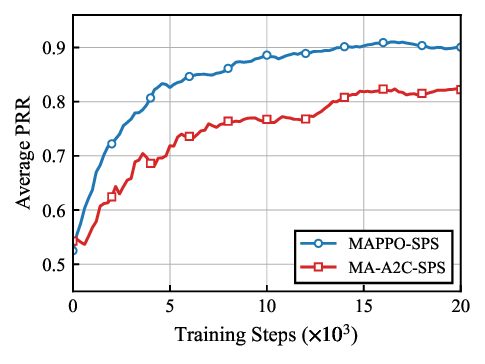}
    \label{fig:marl_prr_training}
}
\hspace{0.045\textwidth}
\subfloat[Range Root-CRLB.]{
    \includegraphics[width=0.34\textwidth]{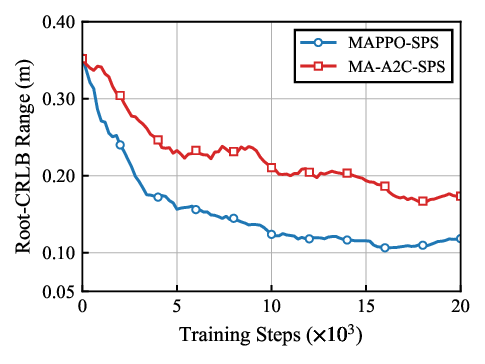}
    \label{fig:marl_range_crlb_training}
}

\vspace{-3mm}

\subfloat[MEC queueing delay.]{
    \includegraphics[width=0.34\textwidth]{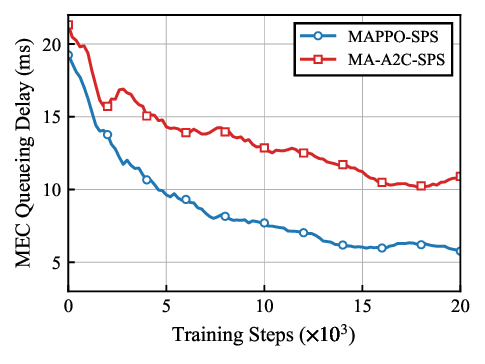}
    \label{fig:marl_mec_queue_training}
}
\hspace{0.045\textwidth}
\subfloat[Normalized reward.]{
    \includegraphics[width=0.34\textwidth]{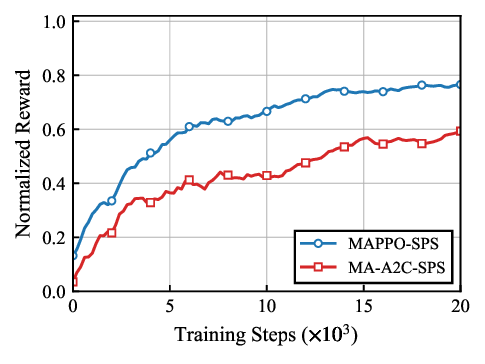}
    \label{fig:marl_reward_training}
}

\caption{Training evolution of communication, sensing, computation, and reward metrics at \(\rho=80\) veh/km.}
\label{fig:marl_training_metrics}
\end{figure*}

In this section, we evaluate \textbf{MAPPO-SPS} against three benchmarks: \textbf{MA-A2C-SPS}, \textbf{SCG-SPS}, and \textbf{CCG-SPS}. The simulation parameters are listed in Table~\ref{tab:iscc_params}, and the comparison covers sensing accuracy, communication performance, and computation efficiency. For the compact KPI comparison in Table~\ref{tab:kpi_compact}, the representative operating points are \(d=80\) m, \(v_{\mathrm{rel}}=20\) m/s, and \(\rho=80\) veh/km. The learning-based schemes were implemented in Python using PyTorch and executed on an NVIDIA GeForce RTX~4080 GPU. In MAPPO-SPS, each vehicle uses a decentralized actor with hidden-layer sizes \(256\), \(128\), and \(64\), while the centralized critic uses hidden-layer sizes \(512\), \(256\), and \(128\) over the aggregated multi-agent state to capture sensing, sidelink, offloading, and MEC-queue coupling during training.

Fig.~\ref{fig:marl_training_metrics} shows the training evolution of MAPPO-SPS and MA-A2C-SPS under dense traffic, \(\rho=80\) veh/km. All curves are averaged over three independent runs with different random seeds. MAPPO-SPS achieves faster PRR improvement and a higher steady-state reliability than MA-A2C-SPS, as shown in Fig.~\ref{fig:marl_prr_training}, indicating stronger sidelink resource adaptation under contention. It also reduces the Root-CRLB for range estimation more rapidly in Fig.~\ref{fig:marl_range_crlb_training}, showing that sensing resources are allocated more effectively without sacrificing communication reliability. In Fig.~\ref{fig:marl_mec_queue_training}, MAPPO-SPS maintains lower MEC queueing delay, reflecting better overflow offloading and queue control. The normalized reward in Fig.~\ref{fig:marl_reward_training} further confirms its higher and more stable convergence. These trends indicate that the centralized critic captures the coupling among sidelink contention, sensing allocation, and computation offloading more effectively. Since SCG-SPS and CCG-SPS are non-learning greedy baselines, they are excluded from this training-evolution comparison and evaluated only in steady state.

\subsection{Sensing Performance Analysis} 
\label{subsec:sensing_results}

\begin{figure*}[!ht]
    \centering
    \begin{minipage}[t]{0.48\linewidth}
        \centering
        \includegraphics[width=7.1cm,height=4.8cm]{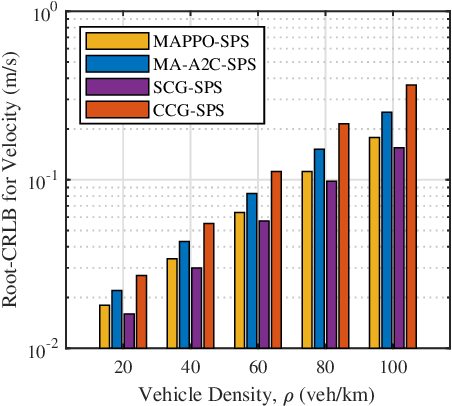}
        \caption{Velocity root-CRLB vs. $\rho$.}
        \label{fig:rootcrlb_vel_density}
    \end{minipage}\hfill
    \begin{minipage}[t]{0.48\linewidth}
        \centering
        \includegraphics[width=7.1cm,height=4.8cm]{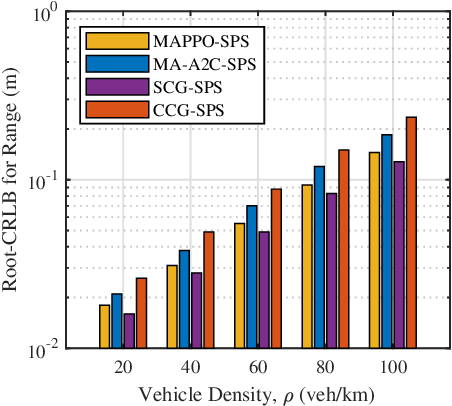}
        \caption{Range root-CRLB vs. $\rho$.}
        \label{fig:rootcrlb_range_density}
    \end{minipage}
\end{figure*}

Figs.~\ref{fig:rootcrlb_vel_density}--\ref{fig:rootcrlb_range_dist} evaluate the sensing performance of MAPPO-SPS and the benchmark schemes. The variation of velocity root-CRLB with vehicle density is presented in Fig.~\ref{fig:rootcrlb_vel_density}. As \(\rho\) increases from \(20\) to \(100\) veh/km, all schemes show degraded Doppler-estimation accuracy because denser traffic increases sidelink contention, raises the probability of resource overlap, and reduces favorable sensing-resource availability. SCG-SPS gives the lowest velocity error since it explicitly prioritizes sensing. MAPPO-SPS remains close to SCG-SPS, reaching about \(1.78\times10^{-1}\) m/s at \(\rho=100\) veh/km, whereas CCG-SPS increases to about \(3.65\times10^{-1}\) m/s. This indicates that MAPPO-SPS retains most of the sensing gain of the greedy sensing-centric scheme while still accounting for communication and computation objectives.

A similar density-dependent degradation is observed for range estimation in Fig.~\ref{fig:rootcrlb_range_density}. The range root-CRLB increases with \(\rho\) due to stronger resource contention and tighter competition among sensing, communication, and offloading PRBs. At \(\rho=100\) veh/km, MAPPO-SPS achieves about \(1.45\times10^{-1}\) m, close to SCG-SPS at \(1.28\times10^{-1}\) m, and lower than MA-A2C-SPS and CCG-SPS. The small gap between SCG-SPS and MAPPO-SPS is expected because SCG-SPS greedily favors sensing, whereas MAPPO-SPS optimizes the joint ISCC objective and therefore avoids over-allocating resources to sensing alone. The distance-dependent range-estimation behavior is illustrated in Fig.~\ref{fig:rootcrlb_range_dist}. The root-CRLB increases with distance because the reflected echo becomes weaker, reducing the sensing SNR term in \eqref{eq:crlb_range}. At short distances, the gap among schemes is limited, but it becomes clearer as distance grows due to the increased sensitivity of echo quality to radio-resource allocation. At \(d=50\) m, MAPPO-SPS achieves a root-CRLB of approximately \(4.9\times10^{-2}\) m, close to SCG-SPS and substantially lower than CCG-SPS. Overall, MAPPO-SPS provides near sensing-centric accuracy while avoiding the single-objective behavior of SCG-SPS.

\subsection{Communication Performance Analysis}
\label{subsec:comm_results}

\begin{figure*}[!ht]
    \centering
    \begin{minipage}[t]{0.48\linewidth}
        \centering
        \includegraphics[width=7.1cm,height=4.8cm]{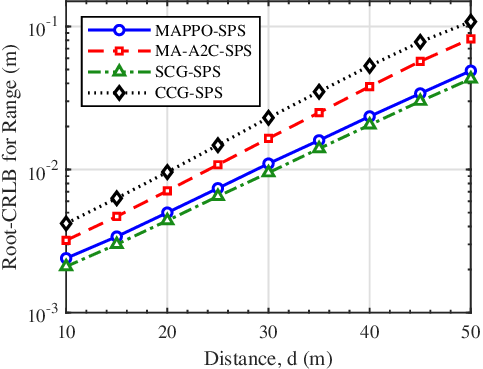}
        \caption{Range root-CRLB vs. distance.}
        \label{fig:rootcrlb_range_dist}
        
    \end{minipage}\hfill
    \begin{minipage}[t]{0.48\linewidth}
        \centering
        \includegraphics[width=7.1cm,height=4.8cm]{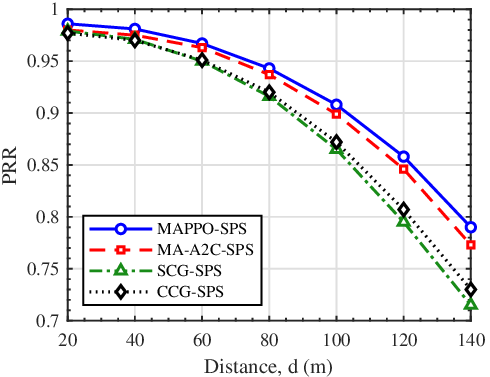}
        \caption{Packet reception ratio versus distance.}
        \label{fig:comm_prr_dist}
        
    \end{minipage}
\end{figure*}

\begin{figure*}[!ht]
    \centering
    \begin{minipage}[t]{0.48\linewidth}
        \centering
        \includegraphics[width=7.1cm,height=4.8cm]{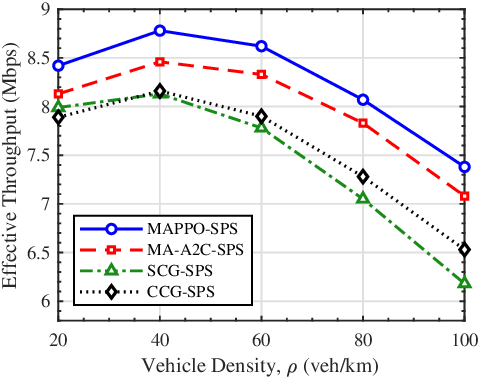}
        \caption{Effective throughput versus vehicle density.}
        \label{fig:comm_throughput_density}
    \end{minipage}\hfill
    \begin{minipage}[t]{0.48\linewidth}
        \centering
        \includegraphics[width=7.1cm,height=4.8cm]{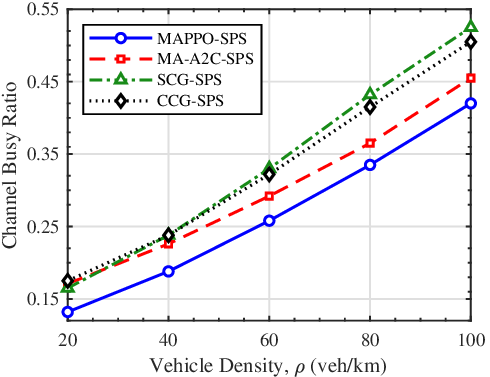}
        \caption{CBR versus vehicle density.}
        \label{fig:comm_cbr_density}
        
    \end{minipage}
\end{figure*}

Fig.~\ref{fig:comm_prr_dist} compares the PRR as the transmitter--receiver distance increases. All schemes maintain high reliability at short distances because the received signal strength is sufficient and the path-loss effect is limited. As distance grows, the PRR decreases due to weaker received power, lower SINR in \eqref{eq:comm_rx_sinr}, and higher decoding failure probability in the receiver-wise PRR model of \eqref{eq:comm_prr}. At \(d=140\) m, MAPPO-SPS maintains a PRR of about \(0.796\), compared with \(0.758\) for MA-A2C-SPS and nearly \(0.72\) for SCG-SPS. This indicates that MAPPO-SPS reduces collision-prone reservations while preserving sufficient communication resources. The effective sidelink throughput under increasing vehicle density is presented in Fig.~\ref{fig:comm_throughput_density}. Throughput initially improves slightly at low-to-moderate density because additional vehicles contribute successful packet deliveries while the resource pool is not yet saturated. Beyond this regime, persistent SB-SPS conflicts and interference dominate, reducing the successfully delivered rate defined in \eqref{eq:comm_effective_bits}. MAPPO-SPS retains about \(7.42\) Mbps at \(\rho=100\) veh/km, compared with \(6.89\) Mbps for MA-A2C-SPS. The lower throughput of SCG-SPS and CCG-SPS follows from their single-objective bias: SCG-SPS reserves more resources for sensing accuracy, whereas CCG-SPS favors computation-side decisions rather than direct sidelink reliability.

The CBR trend in Fig.~\ref{fig:comm_cbr_density} explains the density-dependent throughput loss. As \(\rho\) increases, a larger fraction of the sidelink resource pool is sensed as busy according to \eqref{eq:comm_cbr}, reducing the candidate resource set for SB-SPS selection. At \(\rho=100\) veh/km, MAPPO-SPS keeps the CBR near \(0.42\), while MA-A2C-SPS and the greedy baselines reach higher values. The gap is moderate because the offered load is fundamentally density-driven, but the lower CBR of MAPPO-SPS indicates better avoidance of persistently congested resources. The maximum reliable communication distance is illustrated in Fig.~\ref{fig:comm_max_distance_density}. As vehicle density increases, the reliable range decreases for all schemes because higher CBR and stronger reservation conflicts reduce the probability of maintaining \(\overline{\mathrm{PRR}}(d,\rho)\ge\Gamma_{\mathrm{PRR}}\), as defined in \eqref{eq:max_reliable_distance}. MAPPO-SPS preserves the largest reliable distance across the density range, followed by MA-A2C-SPS. SCG-SPS and CCG-SPS remain close, with CCG-SPS slightly better at higher densities because it is less sensing-biased than SCG-SPS.

\subsection{Computation Performance Analysis}
\label{subsec:comp_results}

\begin{figure*}[!ht]
    \centering
    \begin{minipage}[t]{0.48\linewidth}
        \centering
        \includegraphics[width=7.1cm,height=4.8cm]{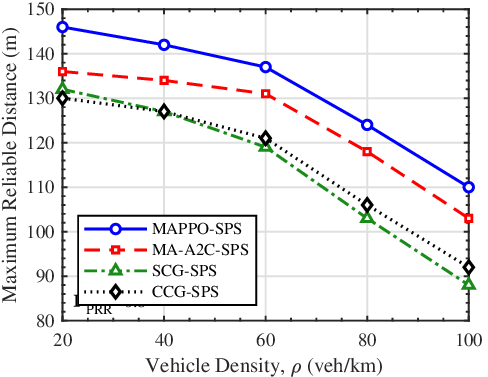}
        \caption{Maximum reliable distance versus vehicle density.}
        \label{fig:comm_max_distance_density}
    \end{minipage}\hfill
    \begin{minipage}[t]{0.48\linewidth}
        \centering
        \includegraphics[width=7.1cm,height=4.8cm]{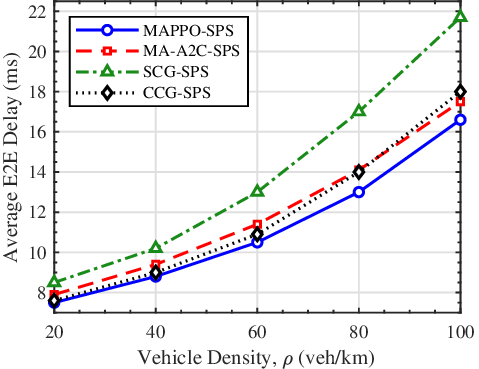}
\caption{Average end-to-end delay versus vehicle density.}
\label{fig:comp_e2e_delay_density}
    \end{minipage}
\end{figure*}

\begin{figure*}[!ht]
    \centering
    \begin{minipage}[t]{0.48\linewidth}
        \centering
        \includegraphics[width=7.1cm,height=4.8cm]{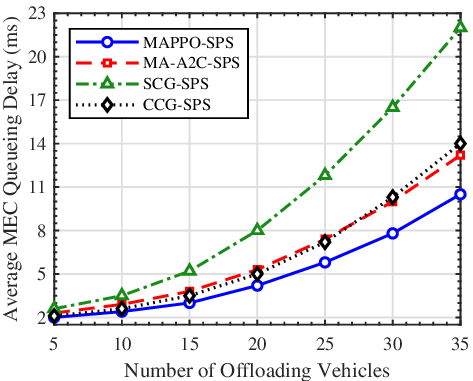}
        \caption{Average MEC queueing delay versus number of offloading vehicles.}
        \label{fig:comp_mec_queue_offloading}
    \end{minipage}\hfill
    \begin{minipage}[t]{0.48\linewidth}
        \centering
        \includegraphics[width=7.1cm,height=4.8cm]{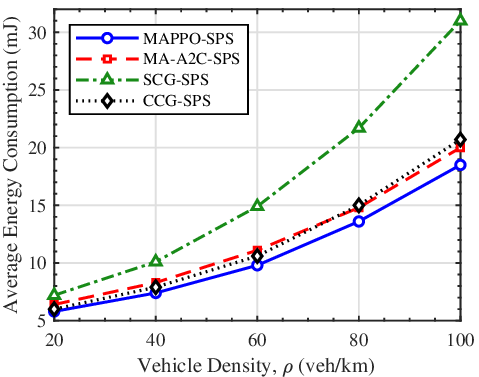}
        \caption{Average vehicle-side energy consumption versus vehicle density.}
        \label{fig:comp_energy_density}
    \end{minipage}
\end{figure*}

The average end-to-end computation--communication delay versus vehicle density is presented in Fig.~\ref{fig:comp_e2e_delay_density}. The delay increases for all schemes as \(\rho\) grows because denser traffic generates larger processing demand. At \(\rho=100\) veh/km, MAPPO-SPS limits the delay to about \(16.6\) ms, compared with \(17.5\) ms for MA-A2C-SPS, \(18.0\) ms for CCG-SPS, and \(21.7\) ms for SCG-SPS. CCG-SPS remains competitive at low and moderate densities since its greedy policy directly minimizes the computation-side cost in \eqref{eq:comp_penalty}. However, its advantage weakens under dense traffic because the myopic rule cannot fully capture future queue buildup and multi-vehicle coupling. The MEC queueing delay against the number of offloading vehicles is illustrated in Fig.~\ref{fig:comp_mec_queue_offloading}. The delay grows gradually under light offloading load, but increases sharply when more vehicles inject overflow workloads into the RSU, reflecting the finite MEC service and queue evolution in \eqref{eq:mec_queue_c_single}--\eqref{eq:mec_queue_s_single}. For \(35\) offloading vehicles, MAPPO-SPS keeps the queueing delay around \(10.5\) ms, whereas MA-A2C-SPS, CCG-SPS, and SCG-SPS reach approximately \(13.2\) ms, \(14.0\) ms, and \(22.0\) ms, respectively. This confirms that MAPPO-SPS regulates offloading according to both local overflow and MEC congestion. CCG-SPS remains strong under low-to-moderate offloading load, but becomes less effective once queue evolution dominates.

Vehicle-side energy consumption under increasing density is compared in Fig.~\ref{fig:comp_energy_density}. Energy consumption rises with density because vehicles perform more local computation, sensing processing, and V2I offloading. This trend is consistent with \eqref{eq:total_vehicle_energy}, which includes local CPU energy, offloading transmission energy, and sensing energy. At \(\rho=100\) veh/km, MAPPO-SPS consumes about \(18.5\) mJ/slot, while MA-A2C-SPS, CCG-SPS, and SCG-SPS consume about \(20.0\), \(20.7\), and \(31.0\) mJ/slot, respectively. SCG-SPS has the highest energy consumption because its CRLB-oriented allocation increases sensing waveform energy and sensing-processing workload. Overall, MAPPO-SPS provides the best latency--queueing--energy tradeoff by balancing local execution, offloading, and sensing-resource usage instead of aggressively favoring a single ISCC function.

\begin{table}[!ht]
\centering
\caption{Performance Comparison of MAPPO-SPS and Benchmark Schemes}
\label{tab:kpi_compact}
\renewcommand{\arraystretch}{1.12}
\setlength{\tabcolsep}{4pt}
\scriptsize
\resizebox{\columnwidth}{!}{%
\begin{tabular}{l c c c c}
\toprule
\textbf{Metric} & \textbf{MAPPO-SPS} & \textbf{MA-A2C-SPS} & \textbf{SCG-SPS} & \textbf{CCG-SPS} \\
\midrule
\multicolumn{5}{c}{\textit{Sensing}} \\
\midrule
Range Root-CRLB (m)              & 0.093 & 0.120 & 0.083 & 0.150 \\
Velocity Root-CRLB (m/s)         & 0.112 & 0.152 & 0.098 & 0.215 \\
\midrule
\multicolumn{5}{c}{\textit{Communication}} \\
\midrule
PRR (\%)                         & 94.3  & 93.7  & 91.6  & 92.0  \\
Effective throughput (Mbps)      & 8.07  & 7.83  & 7.05  & 7.28  \\
CBR (\%)                         & 33.5  & 36.5  & 43.2  & 41.5  \\
Maximum reliable distance (m)    & 124   & 118   & 103   & 106   \\
\midrule
\multicolumn{5}{c}{\textit{Computation}} \\
\midrule
Computation latency (ms)         & 13.0  & 14.1  & 17.0  & 14.0  \\
MEC queueing delay (ms)          & 5.8   & 7.4   & 11.8  & 7.2   \\
Energy consumption (mJ/slot)     & 13.6  & 14.8  & 21.7  & 15.0  \\
\bottomrule
\end{tabular}%
}
\end{table}

Table~\ref{tab:kpi_compact} summarizes the performance indicators at fixed operating points. The sensing, throughput, CBR, computation-latency, and energy metrics are reported at \(\rho=80\) veh/km; PRR is reported at \(d=80\) m; and maximum reliable distance uses \(\Gamma_{\mathrm{PRR}}=0.8\). MEC queueing delay is reported at \(N_{\mathrm{off}}=25\), matching Fig.~\ref{fig:comp_mec_queue_offloading}. MAPPO-SPS gives the best overall ISCC tradeoff, with higher PRR and throughput, lower CBR, larger reliable distance, and reduced computation latency, MEC queueing delay, and energy consumption. SCG-SPS attains the lowest sensing Root-CRLB due to its sensing-centric design, but MAPPO-SPS remains close while substantially improving communication and computation performance. This confirms that MAPPO-SPS balances sensing accuracy, sidelink reliability, and computation efficiency under the same NR-V2X Mode~2 conditions.

\section{Conclusion and Future Work}
\label{sec:conclusion}

In this paper, we proposed an ISCC-aware cross-layer scheduling framework for decentralized NR-V2X Mode~2 networks. We extended SB-SPS by jointly coordinating sensing, sidelink communication, and computation offloading within a unified MAC-layer design. The coupled sensing--communication--computation scheduling problem was modeled as a cooperative Markov game and solved using MAPPO under the CTDE paradigm. Simulation results show that MAPPO-SPS achieves a better cross-layer tradeoff than the benchmarks by improving sensing accuracy, communication reliability, and effective throughput while reducing computation latency and MEC queueing under moderate and dense traffic. These results indicate that MAPPO-SPS does not optimize one ISCC function in isolation, but balances sensing, communication, and computation requirements under decentralized NR-V2X Mode~2 operation. In future work, we will extend the model to multi-RSU deployments, where RSU association, handover-aware offloading, and inter-RSU MEC load balancing will be analyzed.

\bibliographystyle{IEEEtran}
\bibliography{references}

@article{Garcia_CST_2021,
  author  = {M. H. C. Garcia and others},
  title   = {A tutorial on 5G NR V2X communications},
  journal = {IEEE Communications Surveys \& Tutorials},
  volume  = {23},
  number  = {3},
  pages   = {1972--2026},
  month   = jul,
  year    = {2021}
}

@article{Sturm2011_OFDMRadar,
  author  = {C. Sturm and W. Wiesbeck},
  title   = {Waveform Design and Signal Processing Aspects for Fusion of Wireless Communications and Radar Sensing},
  journal = {Proceedings of the IEEE},
  volume  = {99},
  number  = {7},
  pages   = {1236--1259},
  year    = {2011},
  month   = jul
}

@article{Sehla_IoTJ_2022,
  author  = {K. Sehla and T. M. T. Nguyen and G. Pujolle and P. B. Velloso},
  title   = {Resource allocation modes in C-V2X: From LTE-V2X to 5G-V2X},
  journal = {IEEE Internet of Things Journal},
  volume  = {9},
  number  = {11},
  pages   = {8291--8314},
  month   = jun,
  year    = {2022}
}

@article{Todisco_ACCESS_2021,
  author  = {V. Todisco and S. Bartoletti and C. Campolo and A. Molinaro and A. O. Berthet and A. Bazzi},
  title   = {Performance analysis of sidelink 5G-V2X Mode~2 through an open-source simulator},
  journal = {IEEE Access},
  volume  = {9},
  pages   = {145648--145661},
  year    = {2021}
}

@article{Liu_CST_2022,
  author  = {A. Liu and Z. Huang and M. Li and Y. Wan and W. Li and T. X. Han and C. Liu and R. Du and D. K. P. Tan and J. Lu and Y. Shen and F. Colone and K. Chetty},
  title   = {A survey on fundamental limits of integrated sensing and communication},
  journal = {IEEE Communications Surveys \& Tutorials},
  volume  = {24},
  number  = {2},
  pages   = {994--1035},
  year    = {2022}
}

@article{Lu_IoTJ_2024,
  author  = {S. Lu and F. Liu and Y. Li and K. Zhang and H. Huang and J. Zou and X. Li and Y. Dong and F. Dong and J. Zhu and Y. Xiong and W. Yuan and Y. Cui and L. Hanzo},
  title   = {Integrated sensing and communications: Recent advances and ten open challenges},
  journal = {IEEE Internet of Things Journal},
  volume  = {11},
  number  = {11},
  pages   = {19094--19129},
  month   = jun,
  year    = {2024}
}

@article{Long2024DRL,
  author  = {X. Long and Y. Zhao and H. Wu and C.-Z. Xu},
  title   = {Deep reinforcement learning for integrated sensing and communication in RIS-assisted 6G V2X system},
  journal = {IEEE Internet of Things Journal},
  volume  = {11},
  number  = {24},
  pages   = {39834--39849},
  month   = dec,
  year    = {2024}
}

@article{He2023FDISAC,
  author  = {Z. He and others},
  title   = {Full-duplex communication for ISAC: Joint beamforming and power optimization},
  journal = {IEEE Journal on Selected Areas in Communications},
  volume  = {41},
  number  = {9},
  pages   = {2920--2936},
  month   = sep,
  year    = {2023}
}

@article{Wen2024Survey,
  author  = {D. Wen and Y. Zhou and X. Li and Y. Shi and K. Huang and K. B. Letaief},
  title   = {A survey on integrated sensing, communication, and computation},
  journal = {IEEE Communications Surveys \& Tutorials},
  year    = {2024},
  note    = {Early Access, doi: 10.1109/COMST.2024.3521498}
}

@article{Qi2022,
  author  = {Q. Qi and X. Chen and A. Khalili and C. Zhong and Z. Zhang and D. W. K. Ng},
  title   = {Integrating sensing, computing, and communication in 6G wireless networks: Design and optimization},
  journal = {IEEE Transactions on Communications},
  volume  = {70},
  number  = {9},
  pages   = {6212--6227},
  month   = sep,
  year    = {2022}
}

@article{Liang2024,
  author  = {B. Liang and R. Fan and H. Hu and H. Jiang and J. Xu and N. Zhang},
  title   = {Joint task offloading and resource allocation in multi-user mobile edge computing with continuous spectrum sharing},
  journal = {IEEE Transactions on Vehicular Technology},
  volume  = {73},
  number  = {5},
  pages   = {7234--7249},
  month   = may,
  year    = {2024}
}

@article{Tang_CST_2021,
  author    = {F. Tang and Y. Kawamoto and N. Kato and J. Liu},
  title     = {Comprehensive Survey on Machine Learning in Vehicular Network: Technology, Applications and Challenges},
  journal   = {IEEE Communications Surveys \& Tutorials},
  volume    = {23},
  number    = {3},
  pages     = {2027--2057},
  year      = {2021},
  quarter   = {Third Quarter},
  doi       = {10.1109/COMST.2021.3075245}
}

@article{Shao_CommMag_2020,
  author  = {J. Shao and J. Zhang},
  title   = {Communication--Computation Trade-Off in Resource-Constrained Edge Inference},
  journal = {IEEE Communications Magazine},
  volume  = {58},
  number  = {12},
  pages   = {20--26},
  month   = {Dec.},
  year    = {2020},
  doi     = {10.1109/MCOM.001.2000368}
}

@article{Mao2017MEC,
  author  = {Y. Mao and C. You and J. Zhang and K. Huang and K. B. Letaief},
  title   = {A Survey on Mobile Edge Computing: The Communication Perspective},
  journal = {IEEE Communications Surveys \& Tutorials},
  volume  = {19},
  number  = {4},
  pages   = {2322--2358},
  year    = {2017},
  doi     = {10.1109/COMST.2017.2745201}
}

@article{Liu2022UAVISAC,
  author  = {Q. Liu and H. Liang and R. Luo and Q. Liu},
  title   = {Energy-Efficiency Computation Offloading Strategy in UAV Aided V2X Network With Integrated Sensing and Communication},
  journal = {IEEE Open Journal of the Communications Society},
  volume  = {3},
  pages   = {1337--1346},
  year    = {2022},
  doi     = {10.1109/OJCOMS.2022.3195703}
}

@article{liu2023isac_v2x,
  author  = {Q. Liu and R. Luo and H. Liang and Q. Liu},
  title   = {Energy-efficient joint computation offloading and resource allocation strategy for ISAC-aided 6G V2X networks},
  journal = {IEEE Transactions on Green Communications and Networking},
  volume  = {7},
  number  = {1},
  pages   = {413--428},
  month   = mar,
  year    = {2023},
  doi     = {10.1109/TGCN.2023.3240158}
}

@article{Liu2021VECSurvey,
  author  = {Lei Liu and Chen Chen and Qingqi Pei and Sabita Maharjan and Yan Zhang},
  title   = {Vehicular Edge Computing and Networking: A Survey},
  journal = {Mobile Networks and Applications},
  volume  = {26},
  pages   = {1145--1168},
  year    = {2021},
  doi     = {10.1007/s11036-020-01624-1}
}

@article{Tan2022OFDMAEdge,
  author  = {Lin Tan and Zhufang Kuang and Lian Zhao and Anfeng Liu},
  title   = {Energy-Efficient Joint Task Offloading and Resource Allocation in OFDMA-Based Collaborative Edge Computing},
  journal = {IEEE Transactions on Wireless Communications},
  volume  = {21},
  number  = {3},
  pages   = {1960--1972},
  year    = {2022},
  doi     = {10.1109/TWC.2021.3108641}
}

@article{li2025mappo_mec,
  author  = {Li, Han and Xiong, Ke and Lu, Yuping and Chen, Wei and Fan, Pingyi and Letaief, Khaled Ben},
  title   = {Collaborative Task Offloading and Resource Allocation in Small-Cell MEC: A Multi-Agent PPO-Based Scheme},
  journal = {IEEE Transactions on Mobile Computing},
  volume  = {24},
  number  = {3},
  pages   = {2346--2359},
  year    = {2025},
  doi     = {10.1109/TMC.2024.3496536}
}

@article{Capponi_COMST_2019,
  author  = {Andrea Capponi and Claudio Fiandrino and Burak Kantarci and Luca Foschini and Dzmitry Kliazovich and Pascal Bouvry},
  title   = {A Survey on Mobile Crowdsensing Systems: Challenges, Solutions, and Opportunities},
  journal = {IEEE Communications Surveys \& Tutorials},
  volume  = {21},
  number  = {3},
  pages   = {2419--2465},
  year    = {2019},
  doi     = {10.1109/COMST.2019.2914030}
}

@article{Choi_CommMag_2016,
  author  = {J. Choi and V. Va and N. Gonzalez-Prelcic and R. Daniels and C. R. Bhat and R. W. Heath},
  title   = {Millimeter-Wave Vehicular Communication to Support Massive Automotive Sensing},
  journal = {IEEE Communications Magazine},
  volume  = {54},
  number  = {12},
  pages   = {160--167},
  year    = {2016},
  doi     = {10.1109/MCOM.2016.1600071CM}
}

@article{Cai_TNSE_2022,
  author  = {Ting Cai and Zhihua Yang and Yufei Chen and Wuhui Chen and Zibin Zheng and Yang Yu and Hong-Ning Dai},
  title   = {Cooperative Data Sensing and Computation Offloading in UAV-Assisted Crowdsensing With Multi-Agent Deep Reinforcement Learning},
  journal = {IEEE Transactions on Network Science and Engineering},
  volume  = {9},
  number  = {5},
  pages   = {3197--3211},
  year    = {2022},
  doi     = {10.1109/TNSE.2021.3121690}
}

@article{Decarli2024JCSNRV2X,
  author    = {Nicol{\`o} Decarli and Stefania Bartoletti and Alessandro Bazzi and Richard A. Stirling-Gallacher and Barbara M. Masini},
  title     = {Performance Characterization of Joint Communication and Sensing With Beyond 5G NR-V2X Sidelink},
  journal   = {IEEE Transactions on Vehicular Technology},
  volume    = {73},
  number    = {7},
  pages     = {10044--10059},
  year      = {2024},
  doi       = {10.1109/TVT.2024.3365770}
}

@article{Cao_TWC_2020,
  author  = {Xiaowen Cao and Guangxu Zhu and Jie Xu and Kaibin Huang},
  title   = {Optimized Power Control for Over-the-Air Computation in Fading Channels},
  journal = {IEEE Transactions on Wireless Communications},
  volume  = {19},
  number  = {11},
  pages   = {7498--7513},
  year    = {2020},
  doi     = {10.1109/TWC.2020.3012287}
}

@inproceedings{Cao_WiOpt_2018,
  author    = {Xiaowen Cao and Feng Wang and Jie Xu and Rui Zhang and Shuguang Cui},
  title     = {Joint Computation and Communication Cooperation for Mobile Edge Computing},
  booktitle = {Proc. 16th International Symposium on Modeling and Optimization in Mobile, Ad Hoc, and Wireless Networks (WiOpt)},
  pages     = {1--6},
  year      = {2018},
  doi       = {10.23919/WIOPT.2018.8362865}
}

@article{Huang2022ISAC,
  author  = {N. Huang and T. Wang and Y. Wu and Q. Wu and T. Q. S. Quek},
  title   = {Integrated sensing and communication assisted mobile edge computing: An energy-efficient design via intelligent reflecting surface},
  journal = {IEEE Wireless Communications Letters},
  volume  = {11},
  number  = {10},
  pages   = {2085--2089},
  month   = oct,
  year    = {2022}
}

@article{Zhao2022RRAlloc,
  author  = {L. Zhao and D. Wu and L. Zhou and Y. Qian},
  title   = {Radio resource allocation for integrated sensing, communication, and computation networks},
  journal = {IEEE Transactions on Wireless Communications},
  volume  = {21},
  number  = {10},
  pages   = {8675--8687},
  month   = oct,
  year    = {2022}
}

@article{he2024iscc_framework,
  author  = {Y. He and G. Yu and Y. Cai and H. Luo},
  title   = {Integrated sensing, computation, and communication: System framework and performance optimization},
  journal = {IEEE Transactions on Wireless Communications},
  volume  = {23},
  number  = {2},
  pages   = {1114--1128},
  month   = feb,
  year    = {2024}
}

@article{du2024iscc_federated,
  author  = {M. Du and H. Zheng and M. Gao and X. Feng and J. Hu and Y. Chen},
  title   = {Integrated sensing, communication, and computation for over-the-air federated learning in 6G wireless networks},
  journal = {IEEE Internet of Things Journal},
  volume  = {11},
  number  = {21},
  pages   = {35551--35562},
  month   = nov,
  year    = {2024}
}

@article{Liu_TCOM_2025_ISCC,
  author    = {Peng Liu and Xinyi Wang and Zesong Fei and Yuan Wu and Jie Xu and Arumugam Nallanathan},
  title     = {Latency Minimization Oriented Radio and Computation Resource Allocations for 6G V2X Networks With ISCC},
  journal   = {IEEE Transactions on Communications},
  volume    = {73},
  number    = {12},
  pages     = {15851--15865},
  year      = {2025},
  doi       = {10.1109/TCOMM.2025.xxxxx}
}

@article{Sabeeh2023CBR,
  author  = {Sabeeh, S. and Wesolowski, K.},
  title   = {Congestion Control in Autonomous Resource Selection of Cellular-V2X},
  journal = {IEEE Access},
  volume  = {11},
  pages   = {7450--7460},
  year    = {2023}
}

@inproceedings{Althamary_WCMC_2019,
  author    = {I. Althamary and C.-W. Huang and P. Lin},
  title     = {A survey on multi-agent reinforcement learning methods for vehicular networks},
  booktitle = {Proc. 15th Int. Wireless Commun. Mobile Comput. Conf. (IWCMC)},
  year      = {2019},
  pages     = {1154--1159}
}

@article{Gu2022RLSPS,
  author  = {B. Gu and W. Chen and M. Alazab and X. Tan and M. Guizani},
  title   = {Multiagent Reinforcement Learning-Based Semi-Persistent Scheduling Scheme in C-V2X Mode 4},
  journal = {IEEE Transactions on Vehicular Technology},
  volume  = {71},
  number  = {11},
  pages   = {12044--12056},
  year    = {2022},
  doi     = {10.1109/TVT.2022.3189019}
}

@article{Zhou_TVT_2025,
  author  = {Zhou, Xingkai and Hui, Fei and Liu, Jiajia and Wang, Wenbo and Zhang, Junfei},
  title   = {Resource Reservation in C-V2X Networks for Dynamic Traffic Environments: From Vehicle Density-Driven to Deep Reinforcement Learning},
  journal = {IEEE Transactions on Vehicular Technology},
  volume  = {74},
  number  = {11},
  pages   = {17429--17444},
  year    = {2025},
  month   = nov,
  doi     = {10.1109/TVT.2025.3578083}
}

@article{Schulman2017PPO,
  title   = {Proximal Policy Optimization Algorithms},
  author  = {Schulman, John and Wolski, Filip and Dhariwal, Prafulla and Radford, Alec and Klimov, Oleg},
  journal = {arXiv preprint arXiv:1707.06347},
  year    = {2017}
}

@inproceedings{Yu2022MAPPO,
  title     = {The Surprising Effectiveness of PPO in Cooperative, Multi-Agent Games},
  author    = {Yu, Chao and Velu, Akash and Vinitsky, Eugene and Gao, Jiaxuan and Wang, Yu and Bayen, Alexandre and Wu, Yi},
  booktitle = {Advances in Neural Information Processing Systems},
  volume    = {35},
  pages     = {24611--24624},
  year      = {2022}
}

\end{document}